\documentclass{aa}
\usepackage{txfonts}
\usepackage{natbib}
\usepackage{color}
\usepackage{graphicx}
\usepackage{hyperref}

\bibpunct{(}{)}{;}{a}{}{,}

\newcommand{\qz}{\hbox{QZ~Car }}
\newcommand{\qe}{\hbox{QZ~Car}}
\newcommand{\Mnom}{\hbox{$\mathcal{M}^{\mathrm N}_\odot$}}
\newcommand{\Rnom}{\hbox{$\mathcal{R}^{\mathrm N}_\odot$}}

\newcommand{\phoebe}{{\tt PHOEBE} }
\newcommand{\phoebee}{{\tt PHOEBE}}
\newcommand{\spefo}{{\tt SPEFO} }

\newcommand{\fotel}{{\tt FOTEL} }
\newcommand{\fotele}{{\tt FOTEL}}
\newcommand{\korel}{{\tt KOREL} }
\newcommand{\korele}{{\tt KOREL}}
\newcommand{\pyt}{{\tt PYTERPOL} }
\newcommand{\pyte}{{\tt PYTERPOL}}

\newcommand{\AAMM}{\hbox{\AA\,mm$^{-1}$}}

\newcommand{\ubv}{\hbox{$U\!B{}V$}}
\newcommand{\bv}{\hbox{$B\!-\!V$}}
\newcommand{\ub}{\hbox{$U\!-\!B$}}
\newcommand{\bvri}{\hbox{$BV\!RI$}}
\newcommand{\bvr}{\hbox{$BV\!R$}}

\newcommand{\hp}{\hbox{H$_{\rm p}$}}
\newcommand{\oc}{\hbox{$O\!-\!C$}}

\newcommand{\p}{$\pm$}

\newcommand{\m}{$^{\rm m}\!\!.$}

\newcommand{\kms}{km~s$^{-1}$ }
\newcommand{\ks}{km~s$^{-1}$}
\newcommand{\vsin}{$v$~sin~$i$ }

\newcommand{\tef}{$T_{\rm eff}$ }
\newcommand{\teff}{$T_{\rm eff}$}
\newcommand{\lgg}{{\rm log}~$g$ }
\newcommand{\ms}{M$_{\odot}$}
\newcommand{\rs}{R$_{\odot}$}
\newcommand{\cd}{c$\,$d$^{-1}$}
\newcommand{\ha}{H$\alpha$ }

\newcommand{\hae}{H$\alpha$}

\begin{document}

   \title{Towards a consistent model of the hot quadruple system \\
                    HD~93206 = QZ~Carin\ae \ \ $-$\\
            I. Observations and their initial analyses
 \thanks{Based on spectra from observations made with ESO FEROS spectrograph,
Bochum BESO spectrograph and Chiron CTIO spectrograph and on a very rich
collection of photometric observations from many ground-based observing
stations and space photometry from the Hipparcos, BRITE and TESS satellites.}
\thanks{Table 6 is available only in electronic form
at the CDS via anonymous ftp to cdarc.u-strasbg.fr (130.79.128.5)
 or via http://cdsweb.u-strasbg.fr/cgi-bin/qcat?J/A+A/}
}

\titlerunning{Quadruple system HD~93206 = QZ Car}

\author{P. Mayer\thanks{Pavel Mayer passed away
 on the day of his $86^{\rm th}$ birthday Nov. 7, 2018}\inst{1}\and
        P.~Harmanec\inst{1}\and
        P.~Zasche\inst{1}\and
        M.~Bro\v{z}\inst{1}\and
        R.~Catalan-Hurtado\inst{2,15}\and
        B.N.~Barlow\inst{2}\and
        W.~Frondorf\inst{2}\and
        M.~Wolf\inst{1}\and
        H.~Drechsel\inst{3}\and
        R.~Chini\inst{4,5}\and
        A.~Nasseri\inst{4}\and
        A.~Pigulski\inst{6}\and
        J.~Labadie-Bartz\inst{7}\and
        G.W.~Christie\inst{8}\and
        W.S.G.~Walker\inst{9}\and
        M.~Blackford\inst{10}\and
        D.~Blane\inst{11}\and
        A.A.~Henden\inst{12}\and
        T.~Bohlsen\inst{13}\and
        H.~Bo\v{z}i\'c\inst{14}\and
        J.~Jon\'ak\inst{1}}
   \offprints{P. Harmanec,\\
               \email  Petr.Harmanec@mff.cuni.cz}

\institute{
   Astronomical Institute of Charles University,
   Faculty of Mathematics and Physics,\\
   V~Hole\v{s}ovi\v{c}k\'ach~2, CZ-180 00 Praha~8,
   Czech Republic
\and
   Department of Physics, High Point University,
   One University Way, High Point, NC 27268, USA
\and
   Dr.~Karl~Remeis-Observatory \& ECAP, Astronomical Institute,
   Friedrich-Alexander-University Erlangen-N\"urnberg,
   Sternwartstr.~7, 96049 Bamberg, Germany
\and
   Astronomisches Institut, Ruhr-Universit\"at Bochum,
   Universit\"atsstr.~150, 44801 Bochum, Germany
\and
   Instituto de Astronom\'\i a, Universidad Cat\'olica del Norte,
   Avenida Angamos 0610, Antofagasta, Chile
\and
   Instytut Astronomiczny, Uniwersytet Wroc\l awski, Kopernika~11,
   51-622~Wroc\l aw, Poland
\and
    Instituto de Astronomia, Geof\'\i sica e Ciencias Atmosf\'ericas,
    Universidade de S\~ao Paulo, Rua do Mat\~ao 1226, Cidade Universit\'aria,
    05508-900 S\~ao Paulo, SP, Brazil
\and
   Auckland Observatory, PO Box 24180, Royal Oak, Auckland, New Zealand
\and
   Variable Stars South, P O Box 173, Awanui, New Zealand, 0451
\and
   Variable Stars South, Congarinni Observatory, Congarinni, NSW,
   Australia~2447
\and
   Variable Stars South, and Astronomical Society of Southern Africa,
   Henley Observatory, Henley on Klip, Gautenburg, South Africa
\and
   AAVSO, 106 Hawking Pond Road, Center Harbor, NH03226, USA
\and
   Mirranook Observatory, Boorolong Rd Armidale, NSW, 2350, Australia
\and
   Hvar Observatory, Faculty of Geodesy, Zagreb University,
   Ka\v{c}i\'ceva~26, 10000 Zagreb, Croatia
\and
   Eukaryotic Pathogens Innovation Center, Clemson University, Clemson,
   South Carolina 29634, USA
  }

   \date{Release \today}

   \abstract{The hot nine-component system HD~93206, which contains
a~gravitationally bounded eclipsing Ac1+Ac2 binary ($P=5.9987$~d)
and a~spectroscopic Aa1+Aa2 ($P=20.734$~d) binary can provide~important
insights into the origin and evolution of massive stars.
Using archival and new spectra, and a~rich collection of ground-based and space
photometric observations, we carried out a~detailed study of this object.
We provide a much improved description of both short orbits and a good
estimate of the mutual period of both binaries
of about 14500~d (i.e. 40 years). For the first time, we detected weak lines of
the fainter component of the 6.0~d eclipsing binary in the optical region
of the spectrum, measured their radial velocities, and derived
a mass ratio of $M_{\rm Ac2}/M_{\rm Ac1}=1.29$, which is the opposite of what was
estimated from the International Ultraviolet explorer (IUE) spectra.
We confirm that the eclipsing subsystem Ac is semi-detached and is
therefore in a phase of large-scale mass transfer
between its components.
The Roche-lobe filling and spectroscopically brighter component Ac1
is the less massive of the two and is eclipsed in the secondary
minimum. We show that the bulk of the \ha emission, so far believed to be
associated with the eclipsing system, moves with the primary O9.7\,I
component Aa1 of the 20.73~d spectroscopic binary. However, the weak emission in the higher
Balmer lines seems to be associated with the accretion disc
around component Ac2. We demonstrate that accurate masses and other basic physical properties
including the distance of this unique system can be obtained but require
a~more sophisticated modelling. A~first step in this direction is
presented in the accompanying Paper~II (Bro\v{z} et al.).}

   \maketitle
%

\section{Introduction}
The massive multiple system HD~93206 is a very unusual and rare object, which
could potentially be very important for current theories of the origin
of multiple systems and their evolution. If its basic physical properties
can be obtained with sufficient accuracy, we may be able to estimate
the evolutionary age of individual stars and to decide whether they have
a common origin. It is also important to verify whether or not  one of the stars
has already had time to evolve to the stage of evolutionary expansion and
Roche-lobe overflow in the binary subsystem. Our study is a new and rather
detailed contribution to the efforts of a number of previous investigators
to achieve these goals. We collected all available previous individual
observations and obtained numerous new ones. Their analyses are presented here.

\begin{table}
\caption{Magnitudes of the individual components of the multiple
system HD~93206, identified in accordance with the notation
used in the WDS catalogue.}
\label{abcd}
\begin{tabular}{rcccccccc}
\hline\hline\noalign{\smallskip}
Component & $V$  & $H$  & $K$ &Notes \\
          &(mag)&(mag)&(mag)\\
\noalign{\smallskip}\hline\noalign{\smallskip}
        A &6.23  &5.39  &5.25&1   \\
       Aa &      &5.96  &5.85&2   \\
       Ab &10.5  &9.2   &9.6 &3   \\
       Ac &      &6.38  &6.18&2   \\
       Ad &      &      &12.8&4   \\
        B &      &11.3  &10.9&5   \\
        C &14.42 &      &    &6   \\
        D &      &12.8  &12.3&7   \\
\noalign{\smallskip}\hline\noalign{\smallskip}
\end{tabular}
\tablefoot{
1. Integral magnitudes of the whole system A outside binary eclipses
of the Ac1+Ac2 eclipsing binary are $V=6$\m24, $B=6$\m38, and $U=5$\m54
according to our standard Johnson photometry.
The following infrared magnitudes were obtained by
\citet{cutri2003} at JD~2450893.6074, i.e. close to a quadrature of
the Ac1/Ac2 pair (orbital phase 0.815 from the primary minimum):
$J=5$\m551, $H=5$\m393, and $K=5$\m252;
2. \citet{sana2014} resolved component A into very close
components Aa and Ac (called A-D by them), separated by only 0\farcs03.
An accurate position of this pair was then obtained by \citet{sanch2017};
3. At 1\farcs00 from Aa+Ac. Discovered by
\citet{esa97} and later observed by
\citet{toko2010,hart2012,sana2014,toko2015}
and \citet{toko2018};
4. At 0\farcs73 from Aa+Ac. First reported
by \citet{reg2020} and \citet{rainot2020};
5. At an angular distance of 7\farcs07 from A. First noted
by \citet{dawson18} and the latest observation by
\citet{knapp2018} shows no change in its position;
6. At 8\farcs80 from A. Reported by \citet{voute55}
    \citep[we note the correction of the position angle in][]{voute56}
and also observed by \citet{cure2017} and \citet{knapp2018}. The
latter authors identified these components in the Gaia catalogue
and transformed the Gaia magnitude to Johnson $V$;
7. Reported by \citet{sana2014} (and called E by them).
At a distance of 2\farcs58$\pm$0\farcs09 from Aa+Ac.
}
\end{table}

\begin{table*}
\begin{minipage}[]{\textwidth}
\caption{Different notation of individual components of two pairs of
binaries forming the quadruple system \qz used by different investigators.}
\label{key}
\begin{tabular}{l|c|c|c|c|c|c|c|l|c}
\noalign{\smallskip}\hline\noalign{\smallskip}
Brighter component of the 20.7~d SB&Aa1&A&1&A1&Aa&A&Aa\\
Fainter component of the 20.7~d SB &Aa2& &4&A2&Ab&A&Ab\\
Brighter component of the 6~d EB   &Ac1&B&2&B1&Ba&D&B2\\
Fainter component of the 6~d EB    &Ac2& &3&B2&Bb&D&B1\\
\noalign{\smallskip}\hline\noalign{\smallskip}
Used by &1, 2&3&4&5, 7&6 &8&9\\
\hline\noalign{\smallskip}
\end{tabular}
\end{minipage}
\tablefoot{References quoted by numbers in the row ``Used by":
1. \citet{sanch2017};
2. this paper;
3. \citet{MC};
4. \citet{LMS};
5. \citet{MLDA};
6. \citet{stick};
7. \citet{parkin};
8. \citet{sana2014};
9. \citet{walker2017}.
}
\end{table*}

\section{Current knowledge about HD~93206}
\begin{figure}
\centering
\resizebox{\hsize}{!}{\includegraphics{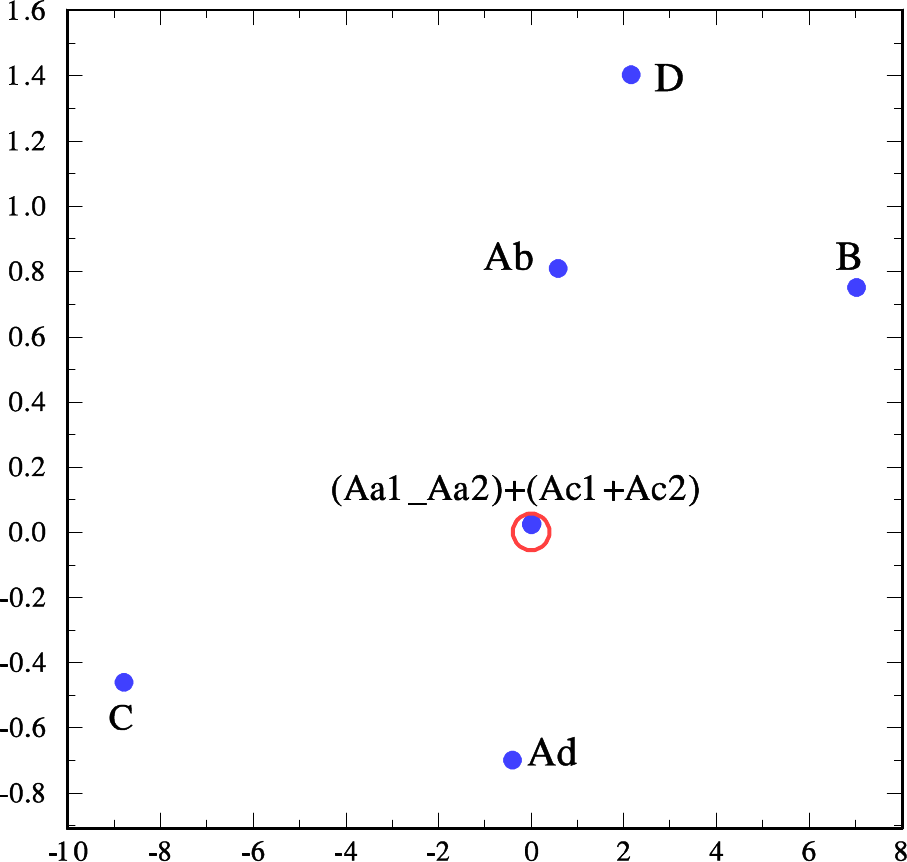}}
\caption{Sketch of the multiple system HD~93206 drawn in arcsec scale
on both axes. We note the expanded scale on the ordinate. The red circle is
centred on the position of the unresolved Aa1-Aa2 spectroscopic binary.
North is up, east is to the left.}
\label{system}
\end{figure}

The massive multiple system known as HD~93206 (HIP~52526), a~member
of the open cluster Collinder~228, consists of at least nine
stars. We adopt for them the same notation as that used
in the Washington Double Stars catalogue (WDS)
\footnote{\url{http://www.astro.gsu.edu/wds/}}.
Their positions in the sky are shown in Fig.~\ref{system}
and their magnitudes, if known, are listed in Table~\ref{abcd}.
It is clear that the flux from the object is dominated by the components
of the quadruple system Aa1+Aa2 and Ac1+Ac2, all other known
components being too faint to affect the photometric
measurements, even if observed with an electronic detector with larger pixel
size. The rest of this study will therefore deal with the close
quadruple system only. A guideline for orientation with the different names of
these four components as used by different investigators is provided in
Table~\ref{key}.

Throughout this paper, we also use the abbreviated form for
heliocentric Julian dates, {RJD=HJD-2400000.0,}
to avoid confusion, which sometimes happens when the half-day
shift introduced by the modified Julian dates is overlooked.

  Periodic light variations of HD~93206 with a 6\fd0 period were discovered
by \citet{walker72} and interpreted as binary eclipses with a $\beta$~Lyr-type 
light curve. The eclipsing binary is now known as QZ~Car.
\citet{conti77} noted the presence of double lines in one photographic
spectrum. Subsequent spectroscopic studies carried out by \citet{MC79},
\citet{LMS}, and \citet{MC} demonstrated that there are two sets of spectral
lines, one showing the radial-velocity (RV) variations with
the eclipsing-binary 6\fd0 period, and the other with a~20\fd7 period,
indicative of the presence of another spectroscopic binary.
\citet{LMS}  also derived the first light-curve solution and concluded
that the eclipsing binary is a semi-detached system. This view was
adopted by \citet{MC}, who also obtained a~series of yellow-red spectra
covering the \ha line. These authors demonstrated the complexity of the observed
spectra and problems of their analyses. Large width of spectral lines in
combination with line blends causes differences in the observed systemic
velocities. In addition, the velocity amplitudes of lower- and higher-ionisation
lines were different.  \citet{MC} also found the presence of a sharp nebular
emission from the Carina nebula at a RV from $-27$ to $-28$~\kms in \ha
and [\ion{O}{ii}]~3729~\AA. They argued that at least part of
the complicated \ha emission moves with the eclipsing Ac system.

\begin{table}
\begin{center}
\caption[]{Published estimates of the masses, radii, and effective
temperatures of the four components of \qe.}\label{masses}.
\begin{tabular}{rccccll}
\hline\hline\noalign{\smallskip}
   Source:    &         1  &       2     &     3\\
\noalign{\smallskip}\hline\noalign{\smallskip}
$M$(Aa1) (\ms)  & (40)       &   $40^*$    & $43\pm6$ \\
$M$(Aa2) (\ms)  &  (9)       &    10       & $19^{+3}_{-7}$ \\
$M$(Ac1) (\ms)  &$16.7\pm5.4$&    14.1     & $20\pm5$ \\
$M$(Ac2) (\ms)  &$28.0\pm7.2$&   $28^*$    & $30\pm5$ \\
\hline
$R$(Aa1) (\rs)  &$22.5\pm2.6$&  $22.5^*$    & 28 \\
$R$(Aa2) (\rs)  &    --      &    6.0      &  6 \\
$R$(Ac1) (\rs)  &$16.1\pm2.4$&   26.9      &$20\pm2$ \\
$R$(Ac2) (\rs)  & $8.9\pm2.4$&  $8.9^*$    &$10\pm2$ \\
\hline
\teff(Aa1) (K)& 32000      & $32000^*$   & 28000 \\
\teff(Aa2) (K)&   --       &  20000      & 33000 \\
\teff(Ac1) (K)& 30000      &  32573      & 30000 \\
\teff(Ac2) (K)& 32463      & $32463^*$   & 36000 \\
\noalign{\smallskip}\hline\noalign{\smallskip}
 \end{tabular}
 \tablefoot{
$^*)$ Adopted from \citet{LMS}. \\
In row "Source", 1: \citet{LMS}, 2: \citet{parkin}, and 3: \citet{walker2017}.
 }
 \end{center}
 \end{table}

\citet{MLDA} obtained new spectral and photometric observations and improved
ephemerides of both orbits. Considering the light-time effect, these authors tentatively
suggested that the orbital period of the Aa-Ac system can be about 50~yr.
\citet{stick} obtained RVs of components Aa1, Ac1, and for the first time
also Ac2 from nine Short Wavelength Prime (SWP) International Ultraviolet Explorer 
(IUE) spectra and derived new orbital elements of both orbits.
For the Ac1-Ac2 subsystem, they obtained a mass ratio
$M_{\rm Ac1}/M_{\rm Ac2}=1.07$, and large values
$M_{\rm Ac1}\sin^3i=43.6\pm3.0$~\ms\ and $M_{\rm Ac2}\sin^3i=40.7\pm2.2$~\ms.

The integral spectrum of the system was classified O9.7\,Ib by
\citet{walborn72} and \citet{sota2014}. \citet{parkin} studied the Chandra
satellite X-ray spectra secured over about 2~years and tried to model them
in terms of colliding winds. These authors found some support for the continuing mass
exchange between the components Ac1 and Ac2. They also revised the component
properties and gave a new spectral classification O9.7\,I for component Aa1,
and O8\,III for component Ac1.
\footnote{In their Table~2, the values of the radius $R$ and mass $M$ of
the component Ac2 (B1 in their notation), taken from \citet{martins}, are
interchanged.}

A detailed study of the system was published by \citet{walker2017}. These authors
derived new RVs and orbital solutions for both orbits and also new photometry
and the light-curve solution. Using these, \citet{walker2017} estimated improved values of
system masses, radii, and effective temperatures. They also concluded that
the period of mutual orbit of the two binaries Aa and Ac is close to 49~years.

\citet{mayer2020} published a conference report on the partial
results of the study presented here. They obtained the first RV curve
of component Ac2 from the spectra in the optical region and, in combination
with photometry, derived the masses of both bodies of the eclipsing subsystem,
finding that Ac2 is the more massive of the two. \citet{mayer2020} also concluded that
the system is a semi-detached one. Analysing variations due to
light-time effect and secular changes of the systemic velocities of both,
Aa and Ac systems, they estimated the orbital period of their mutual
revolution to be about 11700~d (32 years).

\citet{black2020} analysed numerous sets of photometric observations
of \qz including two series of BRITE satellite photometry secured
in 2017 and 2018. They derived epochs of many binary mid-eclipses and
constructed a new \oc \ diagram covering 48 years, using their linear
ephemeris
\begin{equation}
T_{\rm min.I}={\rm JD}~2441033.033+5\fd99857\times E\,.\label{efem-mb}
\end{equation}
\noindent The authors concluded that the Aa-Ac binary had
not yet completed one full orbit since the first set of its photometric
observations in 1971 and noted cycle-to-cycle changes in the shape
of the orbital light curve with the 5\fd999 period.

An overview of the orbital elements published by various investigators
is provided in Appendix~\ref{apt}: In Table~\ref{arvsol}, solutions for
the Aa1-Aa2 20\fd7 spectroscopic binary subsystem are listed; those for the
eclipsing 6\fd0 subsystem Ac1-Ac2 are in Table~\ref{crvsol}.
Several published estimates of the basic physical properties of all four
components are in Table~\ref{masses}.

\section{Observational material used in this study}
\begin{figure}
\centering
\resizebox{\hsize}{!}{\includegraphics{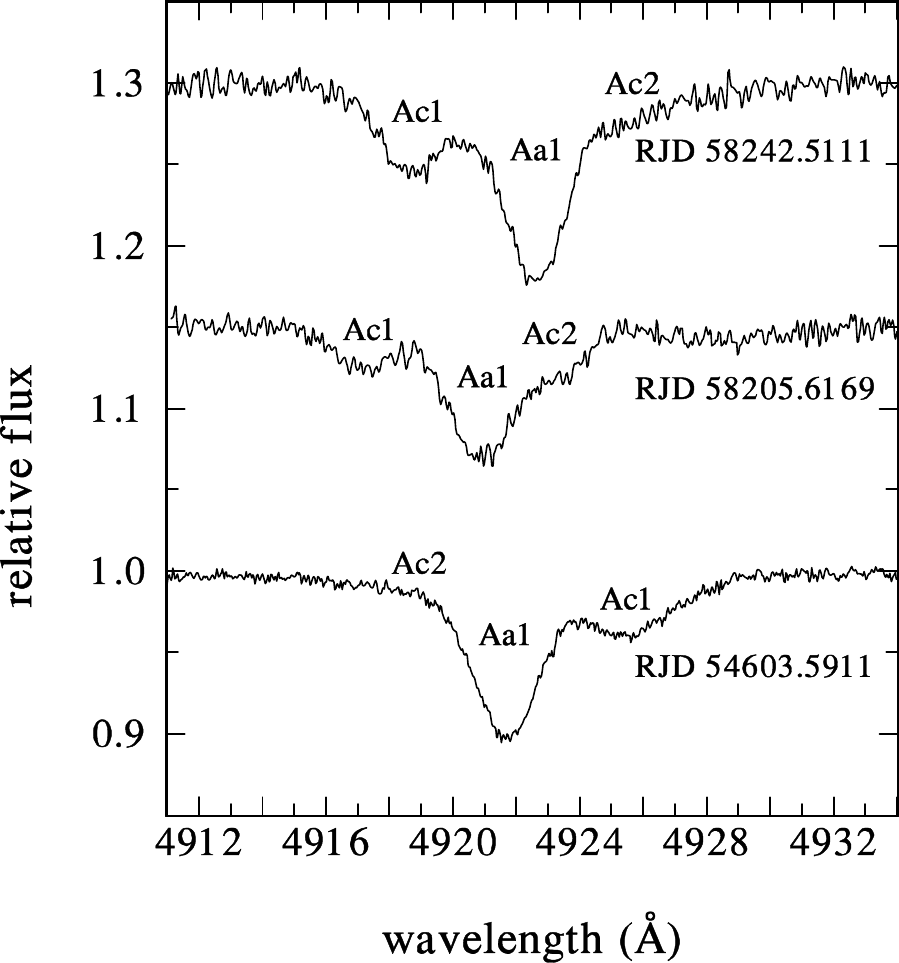}}
\resizebox{\hsize}{!}{\includegraphics{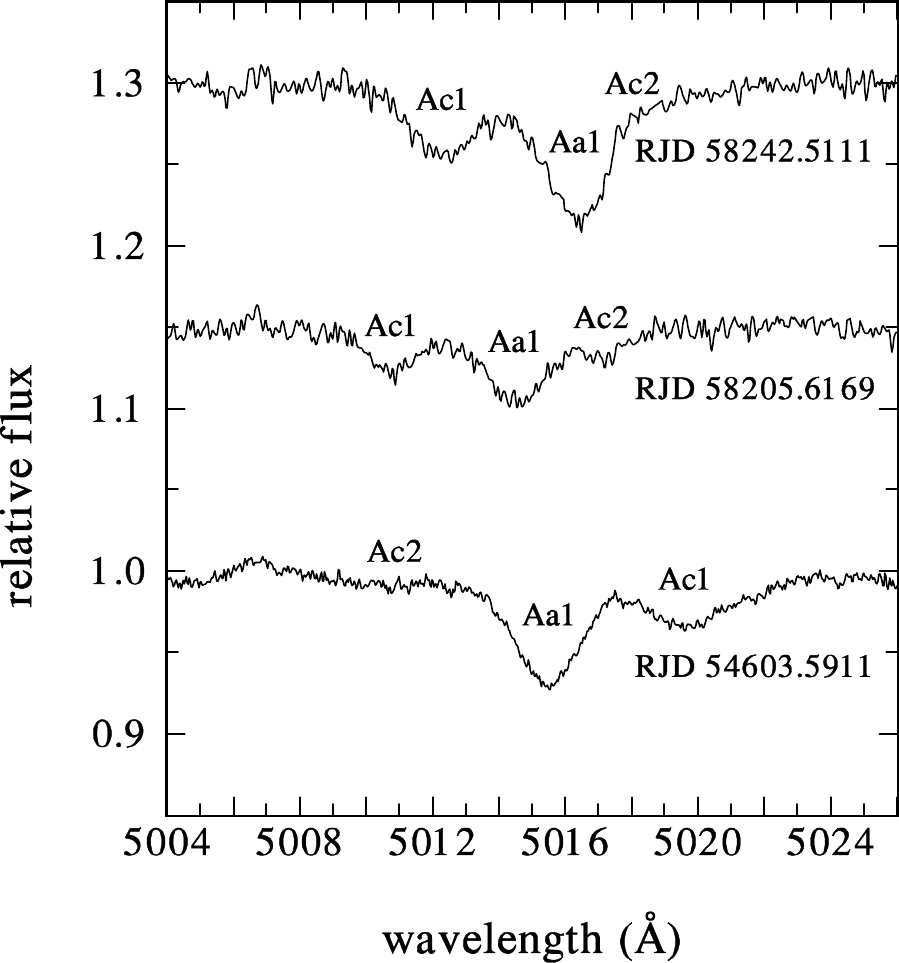}}
\caption{Three examples of the spectra, in which the lines of three components
of the system, Aa1, Ac1, and Ac2 can be measured. Top: \ion{He}{i}~4922~\AA\
line. Bottom: \ion{He}{i}~5016~\AA\ line. RJDs of mid-exposures are shown.
}\label{heprof}
\end{figure}

\subsection{Spectroscopy}
We compiled all published RVs, derived RJDs whenever necessary, and analysed 
79 new electronic spectra from three echelle spectrographs. A~log
of all observations available to us is presented in Table~\ref{jourv}.
Details of the reduction are provided in Appendix~\ref{apa}.
There are some problems with two of our principal data
sets. Echelle spectra from the CTIO Chiron echelle spectrograph have a
very steep blaze function in each echelle order and an accurate spectral
normalisation is not easy, especially in cases where strong spectral
lines are located near the peak of the echelle order or are close to the end
of it. The BESO spectra have low sensitivity, and therefore
a~rather poor signal-to-noise ratio (S/N), and are useful for our purpose only longward
of some 4800~\AA. The best set of available spectra is the collection
of eight FEROS spectra, all of which have a S/N over 200 in the whole available
spectral range.  Their distribution over orbital phases of both close
orbits is rather good. These are the only spectra that can be investigated
in the spectral region from about 4000 to 4400~\AA, where one finds
useful spectral lines sensitive to the radiative parameters of the
system components.
To have a good starting point for the comparison with
published RVs, we measured RVs in all our new spectra using the program
\spefo \citep{sef0,spefo}, namely the latest version 2.63 developed
by J.~Krpata \citep{spefo3}.
It is not easy to find suitable spectral lines for the RV
measurements. Blends, the presence of telluric lines in the red parts
of spectra, and rather strong interstellar lines all complicate the task.
After some trials, we gave up on more sophisticated methods of
RV determination. Guided by previous experience,
we finally selected two stronger singlet \ion{He}{i} lines at 4922 and
5016~\AA, and \ion{He}{ii}~5414~\AA.
In Fig.~\ref{heprof} we show a selection of the \ion{He}{i}~4922~\AA\
and \ion{He}{i}~5016~\AA\ line profiles for some spectra, where a~weak line of
component Ac2 is visible. Whenever possible, we tried to  also measure its RV.

\subsection{Photometry}
We compiled all available accurate photometric observations and tried to reduce
them onto a~comparable system of magnitudes that is close to the Johnson photometric
system. Details of the reduction are provided in Appendix~\ref{apb} and a journal of the
data is shown in Table~\ref{jouphot}. All homogenised ground-based photometric
observations are presented in Table~6, which is available in electronic form only.

\begin{table}
\begin{center}
\caption[]{A log of RV data sets.}\label{jourv}.
\begin{tabular}{ccrccll}
\hline\hline\noalign{\smallskip}
    RJD range  &    No. &  spectral    &   Spg. &  S\\
               &Aa1/Ac1 & region (\AA) & No.\\
\noalign{\smallskip}\hline\noalign{\smallskip}
35594.24--35653.22&  4  &   blue     &  1  &   A \\
36768.52--38125.05&  3  &   blue     &  2  &   B \\
42115.58--43208.78& 7/5 &5400--6800  &  3  &   C \\
42117.73--43240.58&29/17&3300--4900  &  3  &   C \\
43209.65--43534.88&16/16&3600--4900? &  4  &   D \\
43952.98--49813.32& 9/9 &  IUE       &  5  &   E \\
48759.52--49028.67&18/8 &4828--4953  &  6  &   F \\
49147.47--49148.57& 2/1 &4826--5143  &  6  &   F \\
49448.55--49453.59& 5/2 & near 4922  &  7  &   F \\
49452.59--49452.64& 2/0 & near \ha   &  7  &   F \\
53738.82--57157.57& 8/8 &3800--8000  &  8  &   G \\
54941.61--56459.56&17/17&3800--8000  &  9  &   G \\
55879.87--57006.05&34/19&3800--8800  & 10  &   H \\
57163.53--58262.49&54/54&4500--8900   & 11  &   G \\
58867.87--58909.70&27/27&4500--8900& 11 & G \\
 \noalign{\smallskip}\hline\noalign{\smallskip}
 \end{tabular}
 \tablefoot{
Column Spg No identifies telescopes and spectrographs used:
 1. 1.88m Radcliffe reflector, 2-prism Cassegrain sp., 49 A/mm, photographic;
 2. Australian  photographic;
 3. Cerro Tololo 1.5 m reflector, coude and Cassegrain spg., photographic;
 4. ESO 1.5 m reflector coude 12 A/mm spg, IIa-O photographic;
 5. IUE satellite SWP UV spectra;
 6. ESO 1.5 m reflector, ECHELEC spg.;
 7. ESO 1.4 m reflector, CAT/CES spg.;
 8. ESO FEROS spg.;
 9. Bochum Hexapod Telescope, BESO spg.;
10. McKellan 1 m reflector, HERCULES echelle spectrograph, S1600c CCD
11. 1.5 m reflector, CHIRON echelle spg.\\
Column `S' provides references to individal data sets: A. \citet{feast57};
B. \citet{buscombe65};
C. \citet{MC};
D. \citet{LMS};
E. \citet{stick};
F. \citet{MLDA};
G. this paper;
H. \citet{walker2017}.
 }
 \end{center}
 \end{table}

\begin{table*}[th!]
\caption[]{Summary of photometric observations}
\label{jouphot}
\begin{flushleft}
\begin{tabular}{rcccccl}
\hline\hline\noalign{\smallskip}
Source & Station & Epoch & No. of&No. of &HD$_{\rm comp.}$/& Passbands\\
       & No. &HJD-2400000&  obs. &nights &HD$_{\rm check}$ & used     \\
\noalign{\smallskip}\hline\noalign{\smallskip}
A,G&106&41034.92--49433.90&  163& 41& 93131/93695,92740 & \ubv\\
  G&106&41290.07--43270.11&  101& 64& 93131/93695,92740 & \ubv\\
  B&105&42447.81--42484.70&   34& 34& 93222/ --         &$\lambda$ 5170~\AA\\
  C& 61&47885.45--49016.76&  114& 31& space observations& \hp \\
  D& 12&48682.59--49459.56&  293& 20& 93131/92740,93695 & \ubv\\
  G&111&49106.99--49437.10&   29& 18& 93131             & $V$\\
  E&106&49380.90--49422.21&   87& 11& 93695/93131,93737 & \ubv\\
  F& 93&51962.71--55168.84&  682&562& all-sky           & $V$ \\
  G& 11&53116.35--53123.32&   48&  4& 93131/ --         & \ubv\\
  G&109&54987.00--57837.94&  120&109& 93131/93222       & $B{}V$\\
  G& 11&55568.48--55578.58&   30&  5& 93131/93222       & \ubv\\
  G&126&56811.48--57190.59&  504& 27& 92741             & $V$  \\
  G&107&56973.15--57288.25&   55& 52& many              & \bvri\\
H,G&100&57039.00--57124.08&  901& 23& 93131/93695       & \bvr\\
  G&108&57069.22--57173.09&   49& 49& many              & \bvri\\
H,G&100&57441.96--57743.24& 1174& 18& 93131/93695       & \bvr\\
  G&100&57768.96--57808.29&  426&  6& 93131/93695       & \bvr\\
  G&122&57784.56--57806.28&  199& 21& space observations&BRITE $R$\\
H,G&124&57756.46--57966.21&  303&100& 93695             & $V$\\
  G&122&57806.42--57820.31&  133& 14& space observations&BRITE $R$\\
  G&122&57820.37--57843.09&  255& 23& space observations&BRITE $R$\\
H,G&108&57822.98--57823.39&  121&  1& many              & $V$  \\
  G&122&57843.37--57875.92&  394& 33& space observations&BRITE $R$\\
  G&122&57876.00--57936.26&  635& 59& space observations&BRITE $R$\\
  G&123&58165.18--58187.96&  253& 23& space observations&BRITE $R$\\
  G&123&58188.03--58217.97&  424& 30& space observations&BRITE $R$\\
H,G&100&58194.88--58249.13&  599& 10& 93131/93695       & \bvr\\
  G&123&58218.05--58257.48&  468& 40& space observations&BRITE $R$\\
  G&122&58254.47--58258.38&   37&  5& space observations&BRITE $R$\\
  G&122&58258.99--58315.08&  477& 56& space observations&BRITE $R$\\
H,G&124&58512.45--59054.18&  165& 79& 93695             & $V$\\
  G&123&58512.74--58611.26&  413& 39& space observations&BRITE $R$\\
H,G&125&58565.92--58680.97&  986& 17& 93131/93695       & $V$\\
  G&110&58569.47--58595.64&  602& 27& space observations&$\lambda$ 5680--11260~\AA\\
  G&110&58596.83--58623.85&  620& 28& space observations&$\lambda$ 5680--11260~\AA\\
  G&123&58611.33--58681.54&  643& 58& space observations&BRITE $R$\\
  G&122&58887.00--58941.78&  702& 56& space observations&BRITE $R$\\
  G&122&58941.85--59004.86&  638& 58& space observations&BRITE $R$\\
  G&125&59015.87--59074.85&   69& 18& 93131/93695       & $V$ \\
\noalign{\smallskip}\hline\noalign{\smallskip}
\end{tabular}
\tablefoot{
Abbreviations of column ``Source": \\
A... \citet{walker72};
B... \citet{moffat77};
C... \citet{esa97};
D... \citet{mayer92,ma2010} \& this paper;
E... \citet{minima98} and this paper;
F... \citet{pojm2002};
G... this paper;
H... \citet{black2020}.\\
Abbreviations of column ``Stations" (numbers are running
numbers of the observing stations from the Praha/Zagreb data archives):\\
 11... South African Astronomical Observatory 0.5 m reflector Hamamatsu tube;
 12... ESO 0.50 m reflector, La Silla;
 61... Hipparcos $H_p$ magnitude transformed to Johnson $V$ after \cite{hec98};
 93... ASAS3 APT;
100... Chester Hill, Sydney, 0.08~m refractor, Canon 600D DSLR camera;
105... ESO La Silla Bochum 0.61 m reflector, cooled tube;
106... Auckland Observatory 0.50 m Cassegrain reflector, EMI 9502S/A tube;
107... The Bright Star Monitor Australia Station, AstroTech AT-72 refractor,
       Camera: SBIG ST8-XME, owned and operated by Peter Nelson (see
       https://www.aavso.org/bsm-south);
108... Martin Bruce Berry Bright Star Monitor (see https://www.aavso.org/bsm-berry),
       Perth, Australia, currently operated by Greg Bolt;
109... Terrence Bohlsen Australia;
110... TESS broad-band space photometry;
111... Harry Williams Milton Road Observatory;
122... BRITE BHr satellite;
123... BRITE BTr satellite;
124... Henley Observatory, 0.150~m refractor, Canon 1300D DSLR camera.
125... Congarinni Observatory, 0.08 m refractor stopped down to 0.05 m;
126... FRAM 0.30~m automatic monitor, Pierre Auger Observatory, Argentina.
}
\end{flushleft}
\end{table*}

\section{Subsequent data analysis}

\begin{figure}
\centering
\resizebox{\hsize}{!}{\includegraphics{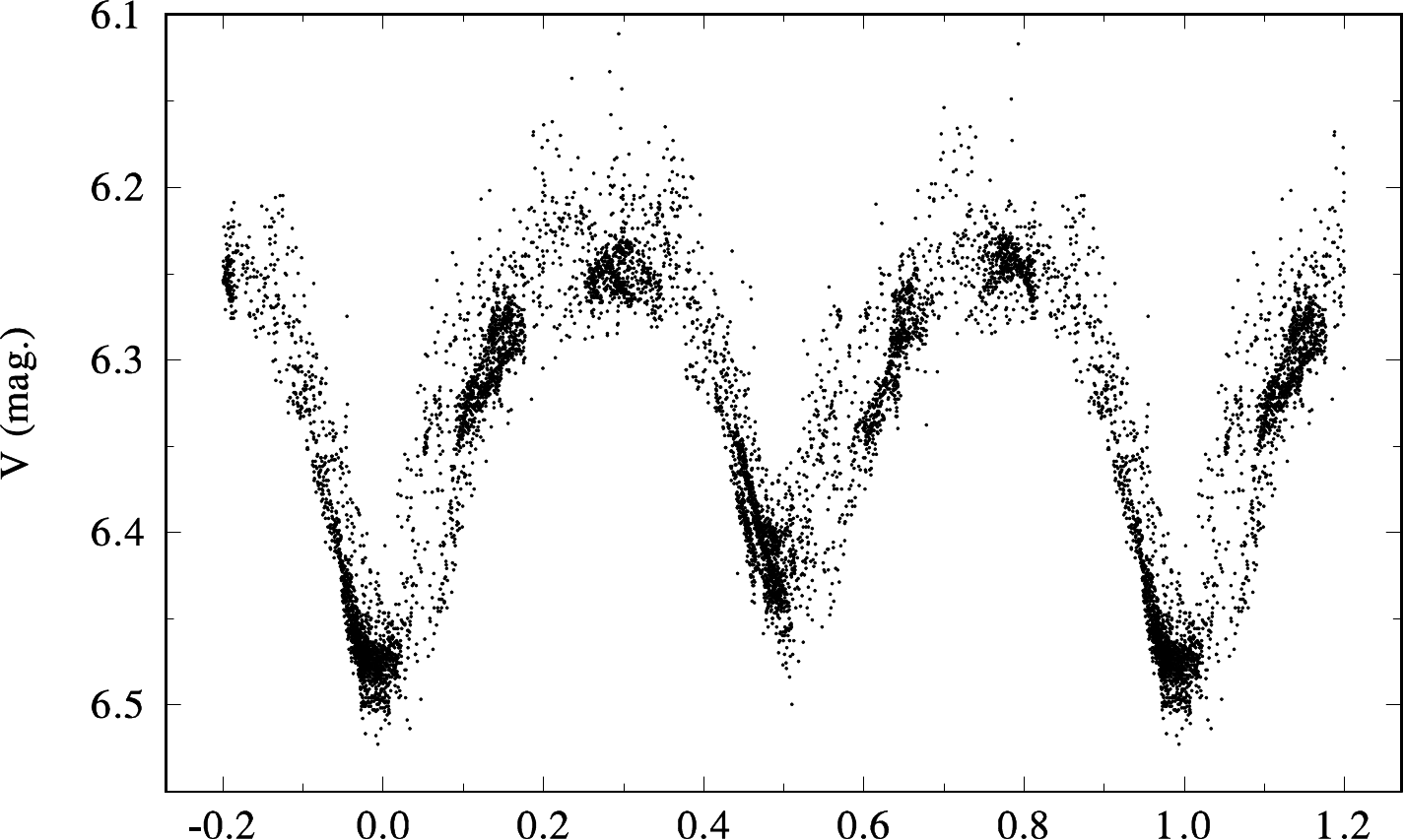}}
\resizebox{\hsize}{!}{\includegraphics{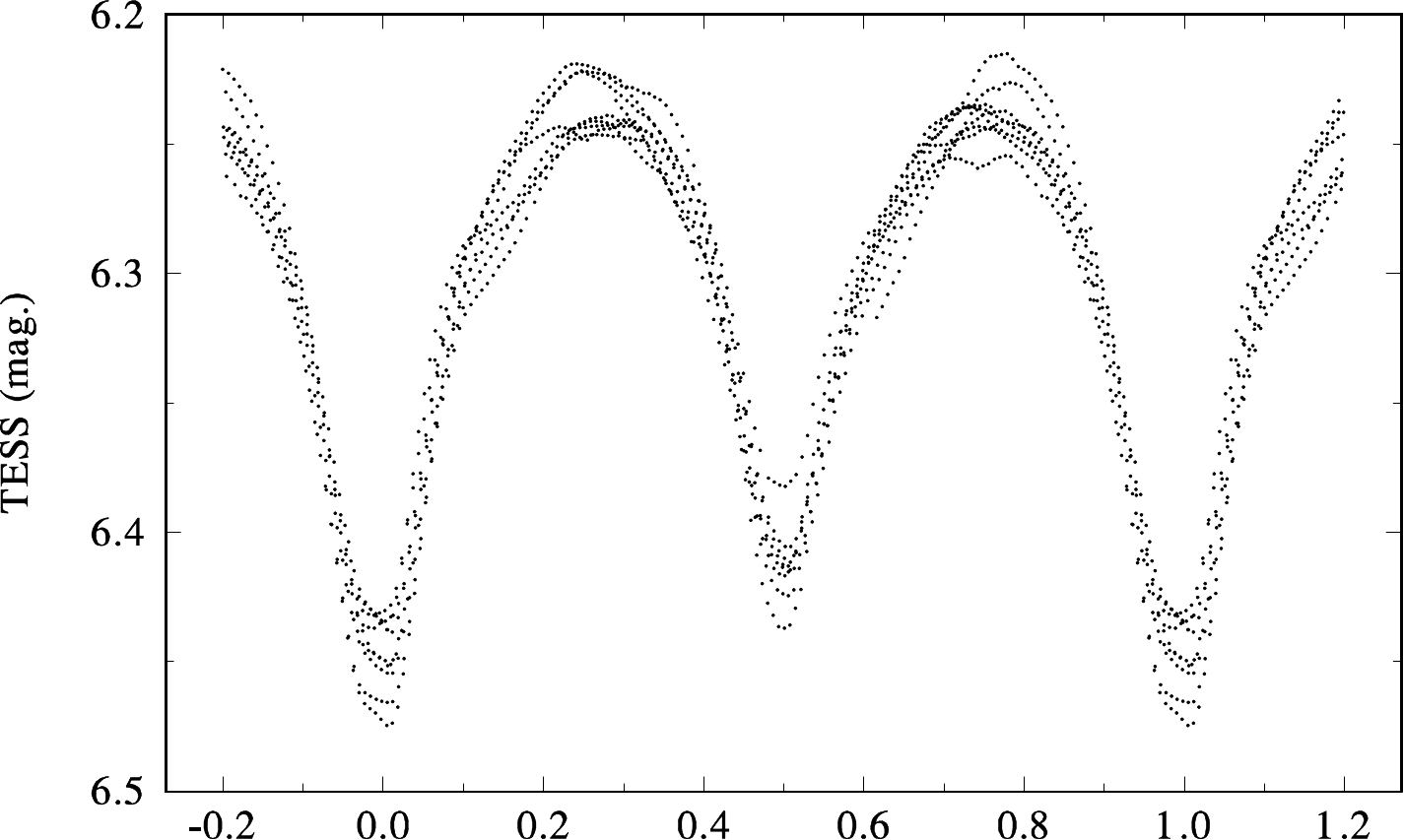}}
\resizebox{\hsize}{!}{\includegraphics{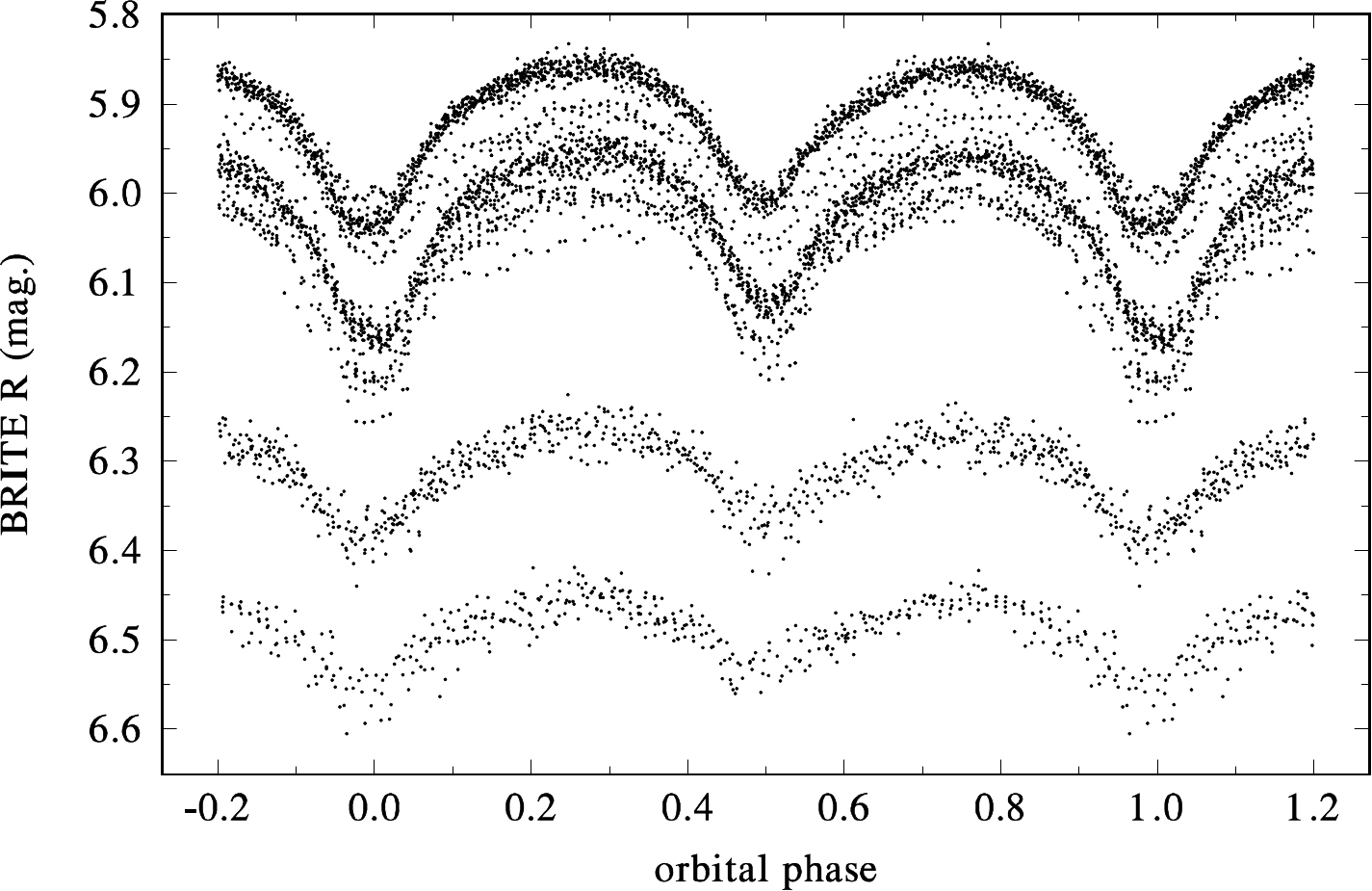}}
\caption{Light curve of the eclipsing system Ac based on all homogenised $V$
band observations at our disposal, and on all BRITE $R$ band and
TESS broad band space photometries plotted for the ephemeris (\ref{efphoI}).
We deliberately show the BRITE data for their original `setups', which have
different zero levels (see Appendix \ref{apb} for details). We note that
the bottom two BRITE light curves have lower amplitudes than the rest.
This is caused by the fact that part of the stellar light was captured by
pixels in the defective part of the detector. These light curves are not
usable for the final light-curve solution but can still be used to make local
determinations of the times of minima discussed here.}
\label{allv}
\end{figure}
In spite of an~extensive set of observational data, the analysis of
a~multiple star system as complicated as \qz is difficult and one
has to proceed in consecutive steps.
We first used all homogenised photometric data in the $V$ band and all
available BRITE and TESS space photometries to derive the most accurate mean
orbital period of system Ac and the epoch of the primary minimum.
A~formal solution of all these observations in the program \phoebe1
\citep{prsa2005,prsa2006} led to the following photometric mean ephemerides:

\begin{eqnarray}
T_{\rm min.I}&=&{\rm RJD}~49425.1096(29)+5\fd998682(2)\times E\,,\label{efphoI}\\
T_{\rm min.II}&=&{\rm RJD}~49428.1089(29)+5\fd998682(2)\times E\,.\label{efphoII}
\end{eqnarray}

\noindent The phase plot of all $V$ data, TESS, and BRITE
photometries for this ephemeris is shown in Fig.~\ref{allv}. The $V$ band
data spanning the longest time interval clearly show
that there are minor changes in the times of minima, which are related
to the light travel time effect in the wide orbit, as we show below.
Additionally, there are obvious cycle-to-cycle changes in the light-curve
shape, which cannot (only) be due to possible microvariability of the comparison
stars used, because they are clearly seen in all space photometries, as shown in
Fig.~\ref{allv} . These phenomena inevitably affect the accuracy
with which the times of minima and the elements of the light-curve solution
are derived.

We then proceeded to new orbital solutions based on all available RVs
in order to obtain improved ephemerides for both orbits.
For all new orbital solutions presented here, we used the program \fotel
\citep{fotel1,fotel2}.

\begin{figure*}[t]
\centering
\includegraphics[angle=0,scale=0.40]{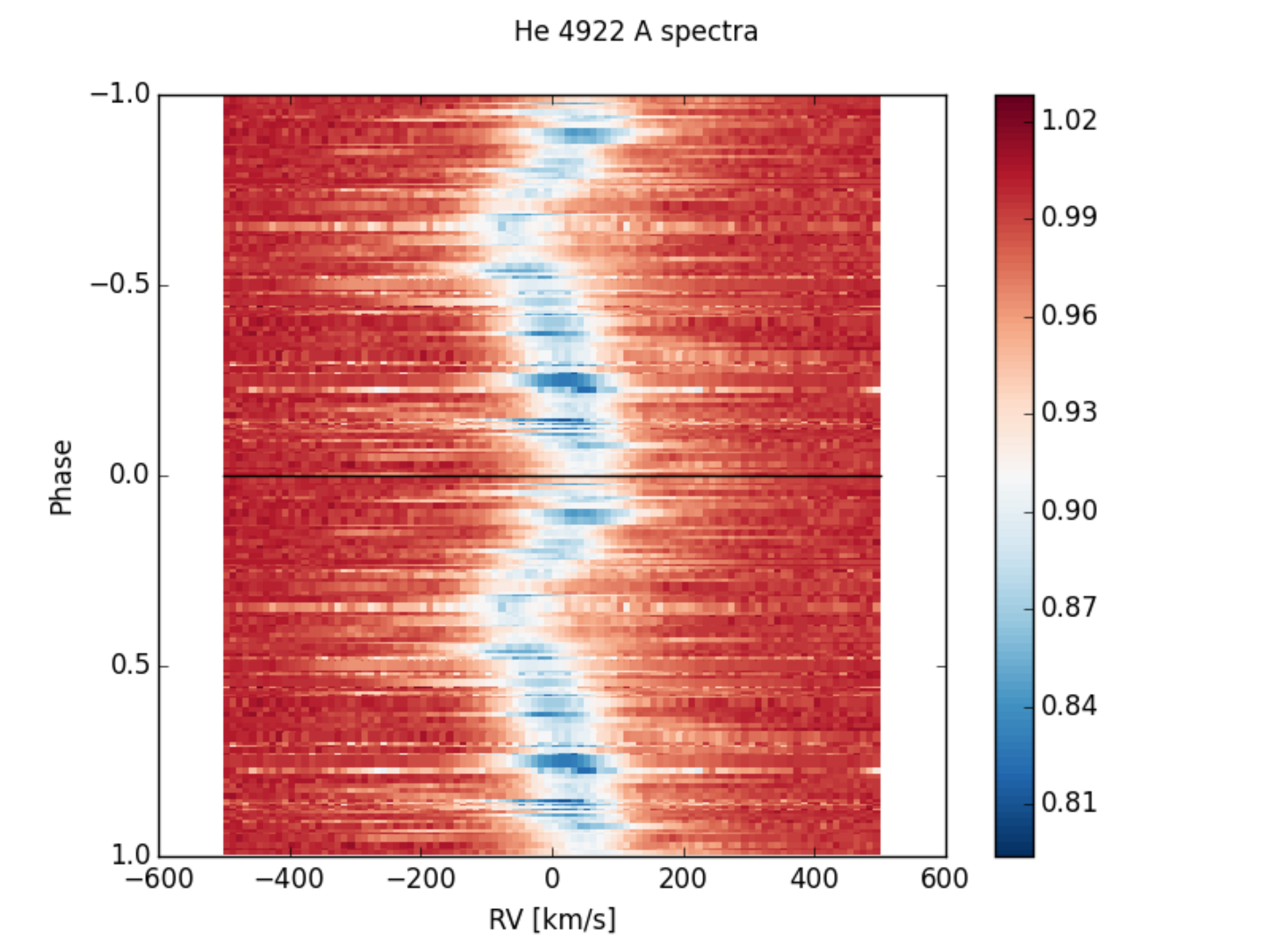}
\includegraphics[angle=0,scale=0.40]{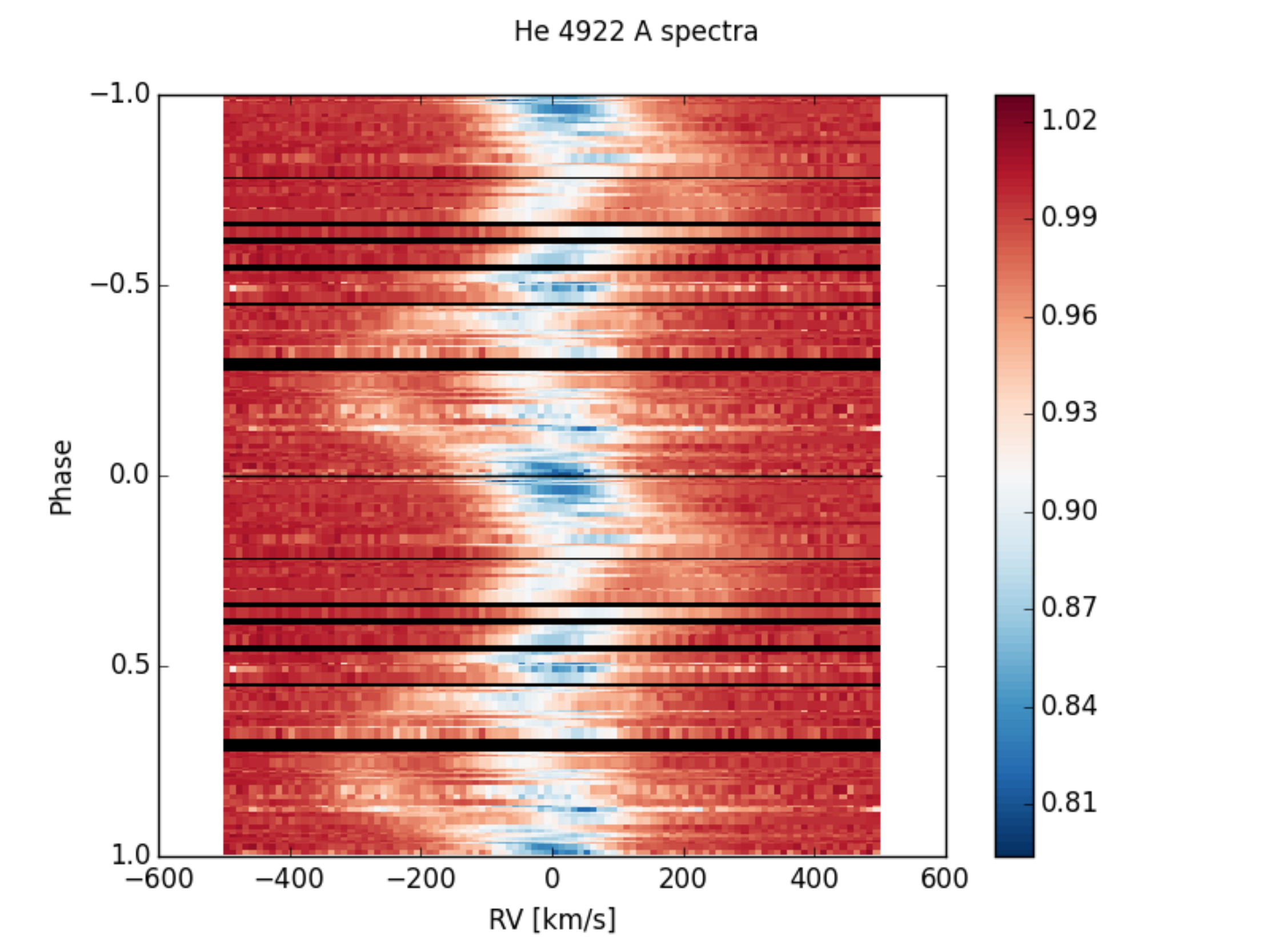}
\includegraphics[angle=0,scale=0.40]{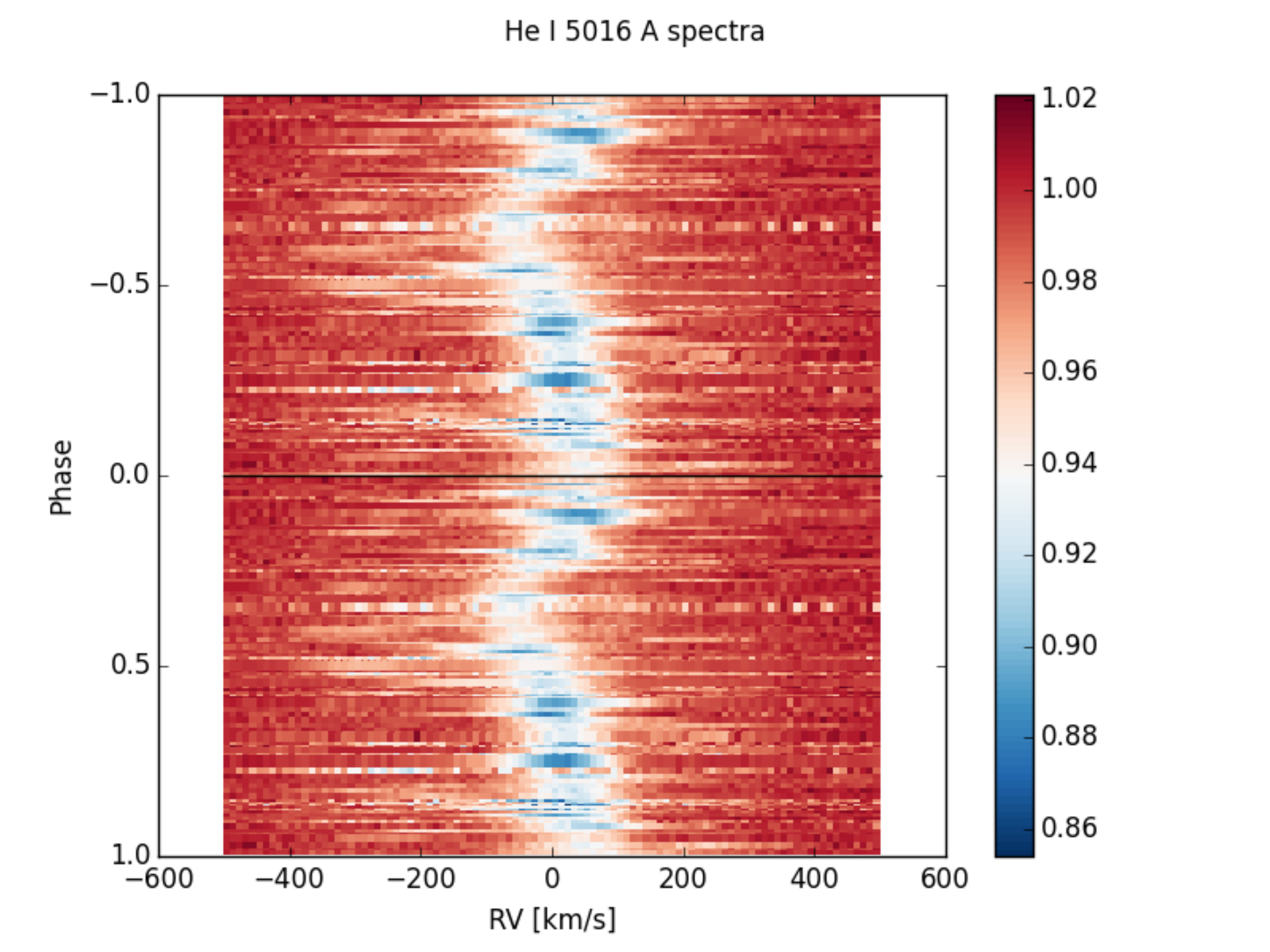}
\includegraphics[angle=0,scale=0.40]{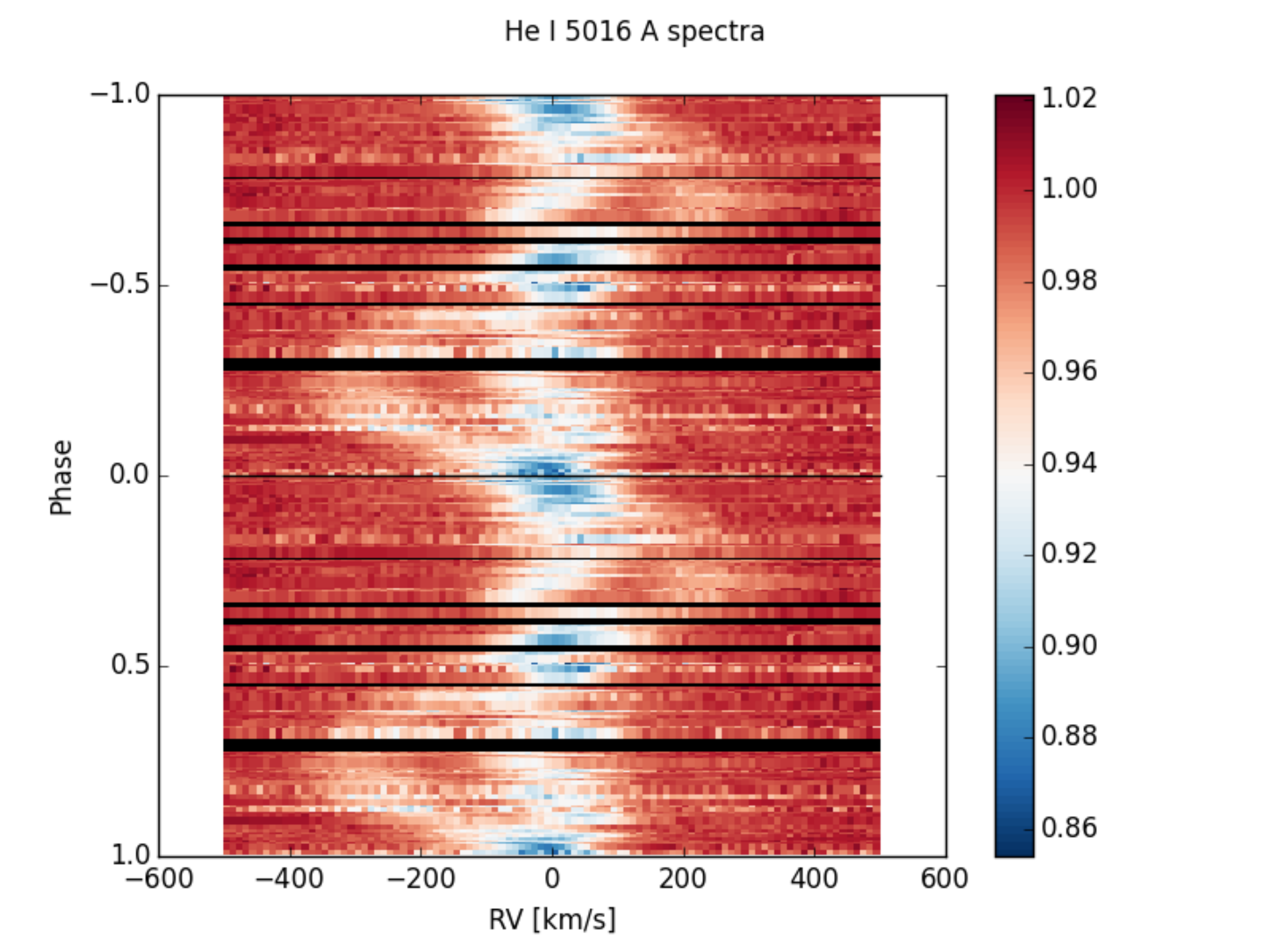}
\includegraphics[angle=0,scale=0.40]{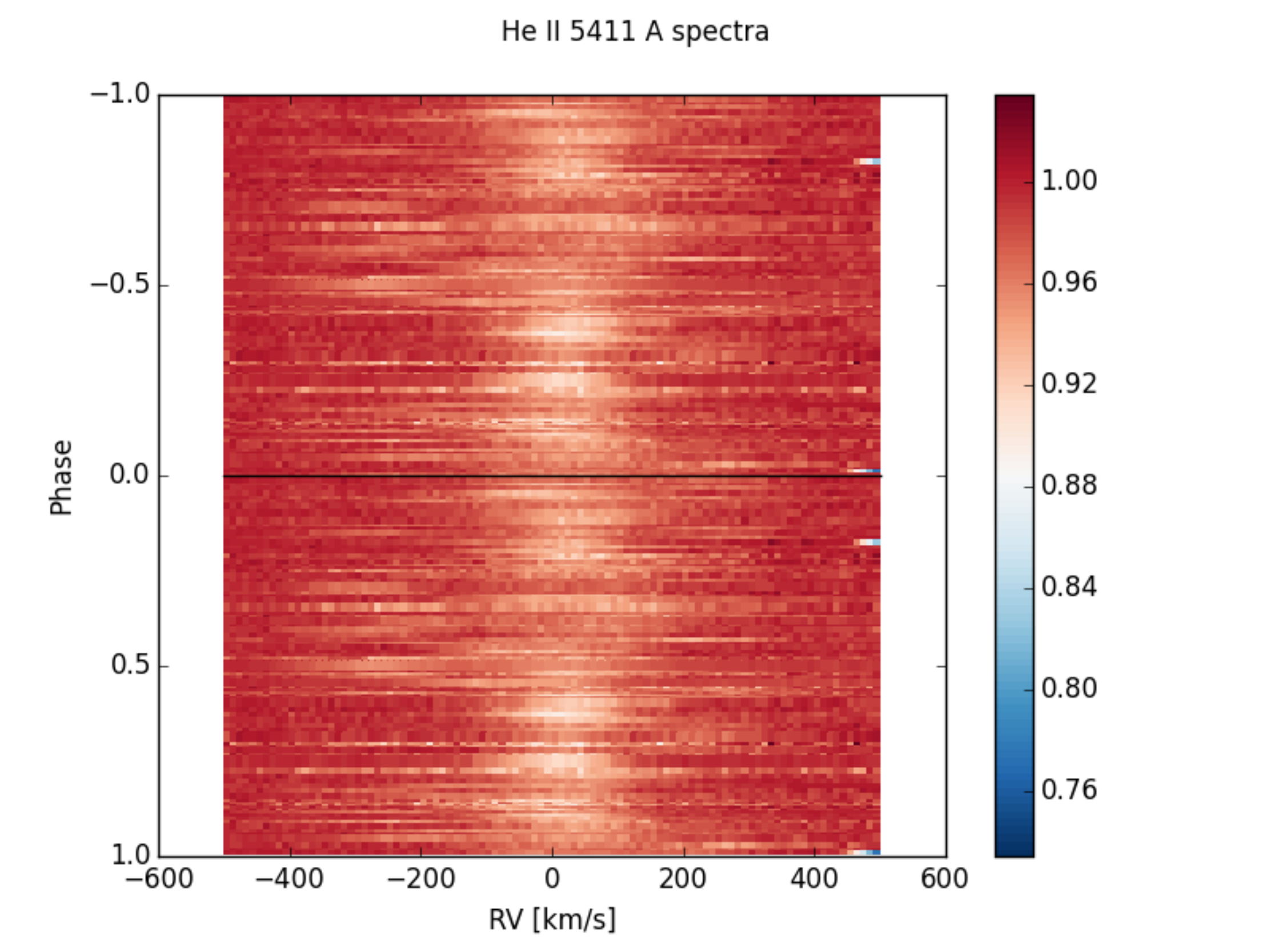}
\includegraphics[angle=0,scale=0.40]{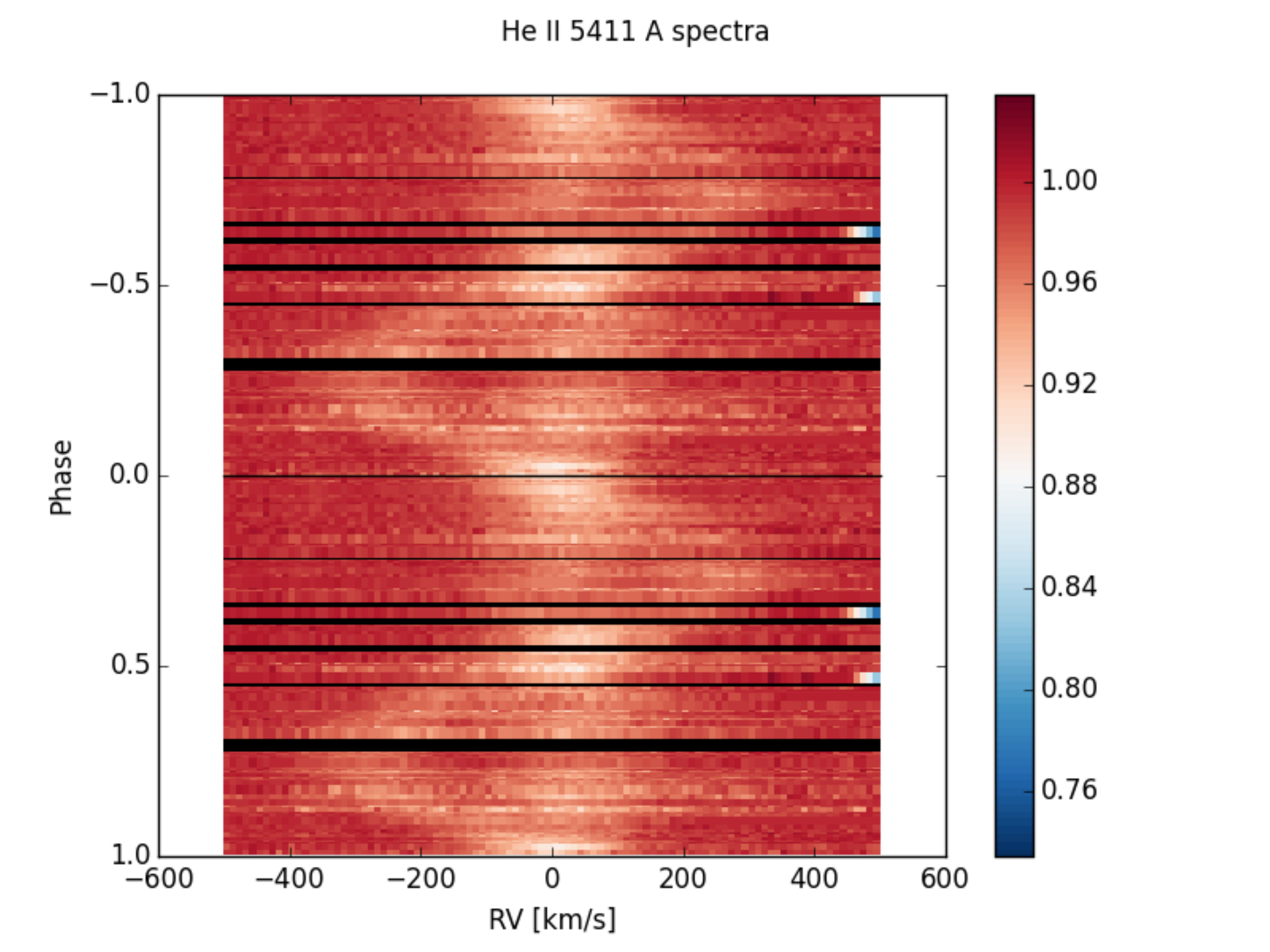}
\caption{Trailed spectra of all FEROS, BESO, and CHIRON line profiles
of the three He spectral lines versus orbital phase, for the 20\fd73419 period (left), with phase zero at RV maximum from solution~C
of Table~\ref{solaa}, and for the 5\fd998682 period (right), with phase zero
from ephemeris (\ref{efphoI}).}\label{trailed}
\end{figure*}

\subsection{Orbit of the spectroscopic component Aa1}
To derive an~updated and improved solution for component Aa1
we used all compiled and new \spefo RVs. Inspection of \spefo RVs
versus phase of the 20\fd73 period led to a surprising finding. While the RVs
for the \ion{He}{i}~4922 and 5016~\AA\ both define the orbital changes
quite well, the RVs from the \ion{He}{ii}~5411~\AA\ line showed no
dependence on the phase of the 20\fd73 period. The mean RV of this line
is about $+20$~\ks, significantly more positive than the systemic velocity
of the whole system. This is clearly seen in the trailed spectra created with
a Python program of J.~Jon\'ak and shown in
the left panels of Fig.~\ref{trailed}. We therefore adopted the mean RVs
of the two \ion{He}{i} lines only to study the 20\fd73 orbital changes.

\begin{figure}
\centering
\resizebox{\hsize}{!}{\includegraphics{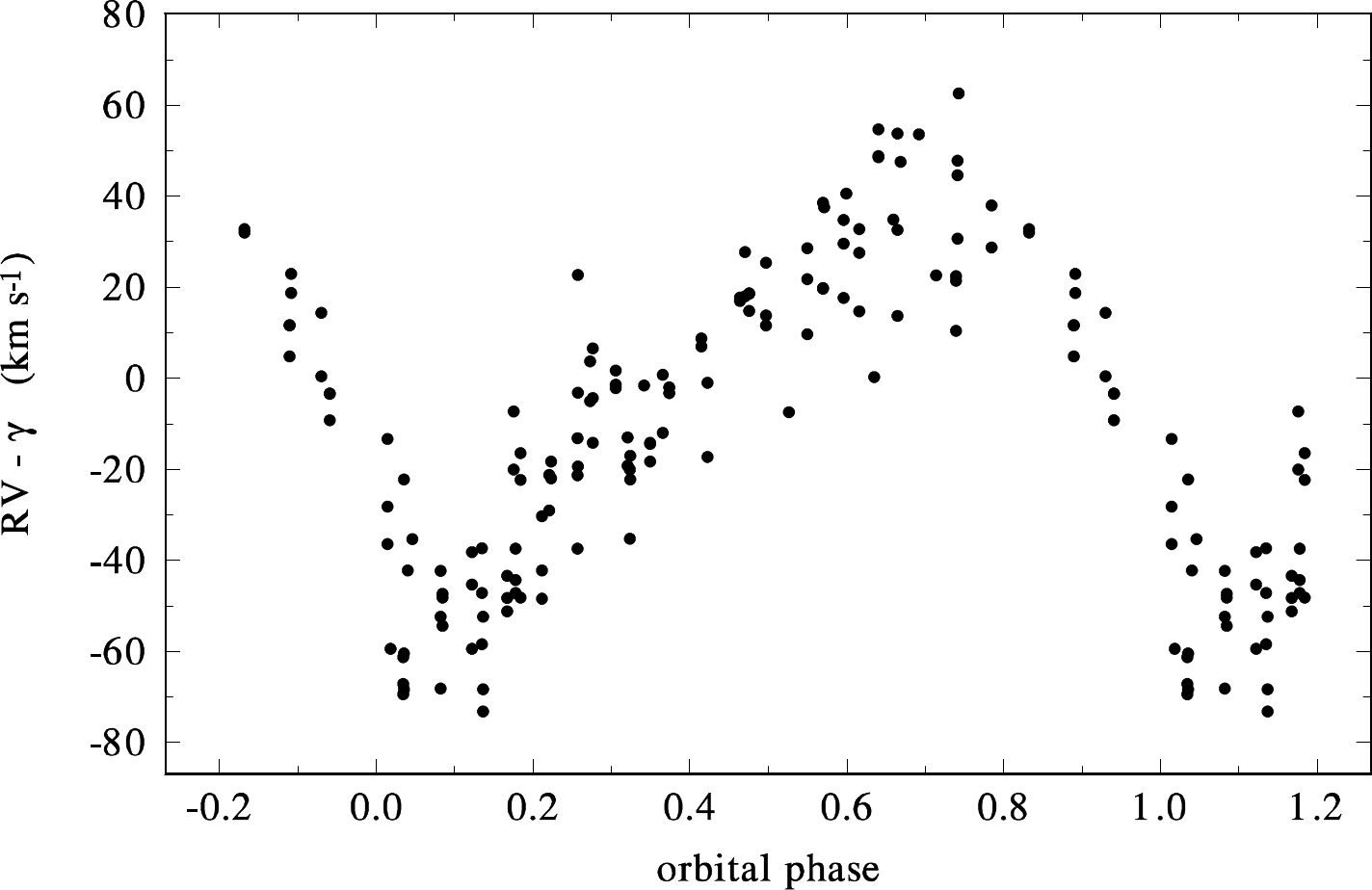}}
\resizebox{\hsize}{!}{\includegraphics{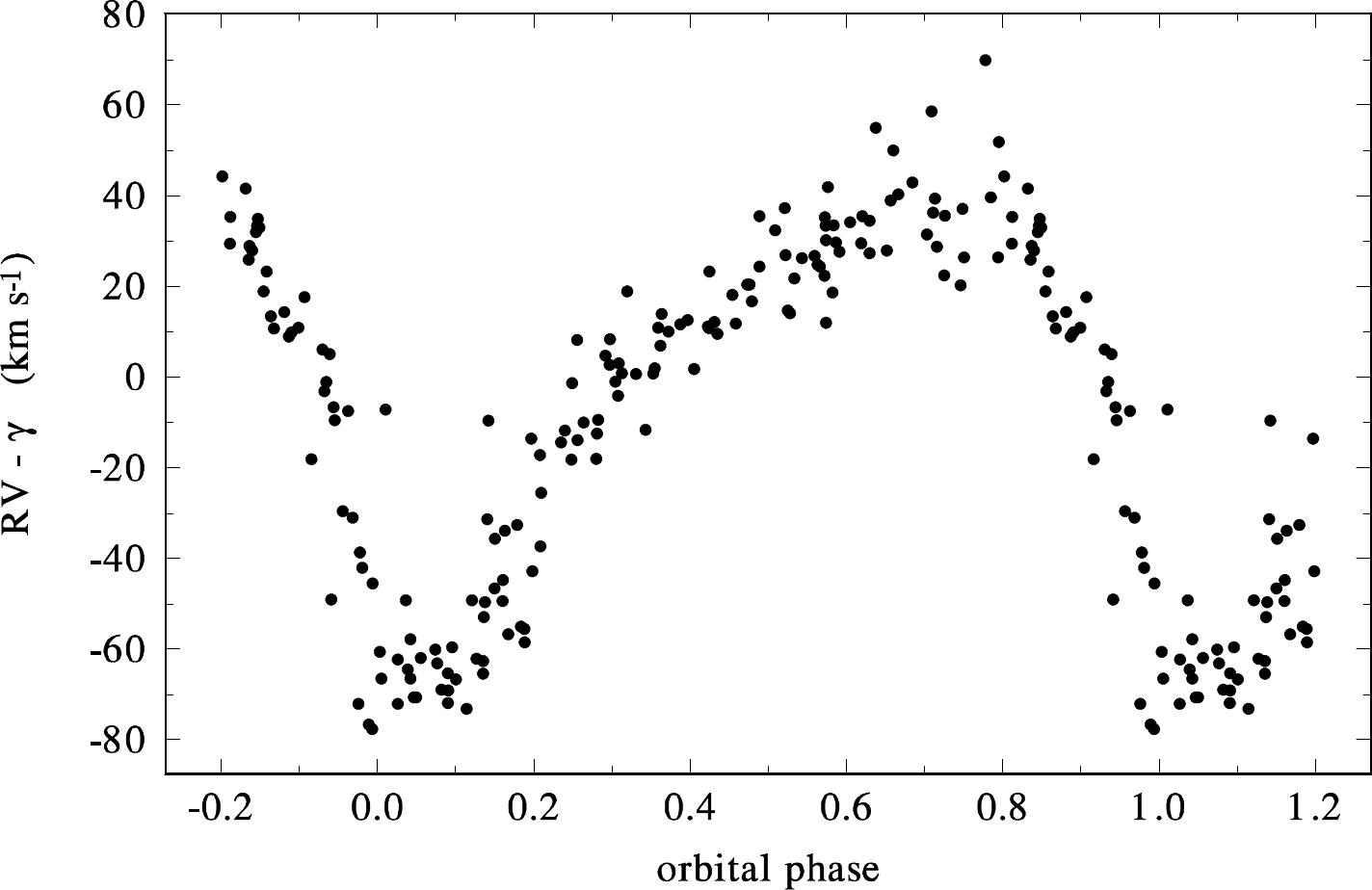}}
\caption{Radial velocity curve of component Aa1 plotted for the period
of 20\fd73419. Top: Older data prior to RJD~45000. Bottom: More recent
data for RJDs after 48000.}
\label{rvc-aa1}
\end{figure}

In order to also monitor the slow change
in the systemic velocity as the subsystem Aa moves in a~wide orbit with
the pair Ac, we sorted the data in time and assigned individual $\gamma$
velocities to data subsets for time intervals not longer than some 1100~d
and much shorter in most cases. Our solution is shown
in Table~\ref{solaa} and the corresponding RV curve is in Fig.~\ref{rvc-aa1}.
Individual local $\gamma$ velocities, together with their mean RJDs,
standard errors, number of individual RVs and time interval covered,
are provided in Table~\ref{gam-aa1}. For photographic spectra from
spectrographs 3 and 4 (see Table~\ref{jourv}), we used the published
measurements for different groups of lines and treated them as
different spectrographs. This confirmed  the existence of some significant differences
in the RV zero point between different groups of lines, as already
noted by previous investigators.

\setcounter{table}{6}
\begin{table}
\begin{flushleft}
\caption{New orbital solutions for RVs of component Aa1 with
locally derived systemic ($\gamma$) velocities.}
\label{solaa}
\begin{tabular}{lrrrrrrrrccc}
\hline\hline\noalign{\smallskip}
 Element           &Solution A  & Solution B & Solution C\\
\noalign{\smallskip}\hline\noalign{\smallskip}
$P$ (d)            &20.73419(18)& fixed      &fixed       \\
$T_{\rm periastr.}$&42530.32(20)&42529.88(40)&53830.59(19)\\
$T_{\rm inf.conj.}$&42529.23    &42528.97    &53829.40    \\
$T_{\rm max.RV}$   &42524.85    &42524.95    &53824.83    \\
 $e$               &0.390(21)   &0.371(42)   &0.387(23)   \\
$\omega$ ($^\circ$)&134.5(3.9)  &125.9(8.4)  &137.5(4.5)  \\
$K_1$              & 50.6(1.6)  &43.4(3.0)   &53.4(1.6)   \\
No. of RVs         & 301        & 134        &167         \\
rms                & 12.48      & 12.96      &11.75 \\
\hline\noalign{\smallskip}
\end{tabular}
\end{flushleft}
\tablefoot{All epochs are in RJDs. In column `rms' we provide the rms
of 1~observation. This, as well as the semi-amplitude $K_1$ , are in \ks.
Solution~A is for all available RVs listed in Table~\ref{jourv}.
Solutions~B and C are for data subsets, with
the period fixed from Solution~A. Solution~B
is for older RVs before RJD~44000, while solution~C is for more recent
electronic spectra. For our new RVs measured with \spefo RVs, we used
mean values of the \ion{He}{i}~4922~\AA, and \ion{He}{i}~5016~\AA\ lines.
This subset also includes RVs obtained by \citet{walker2017}
from high-resolution spectra.}
\end{table}

  Splitting the data into two subsets and calculating trial orbital solutions
for them, we verified that there is currently no compelling evidence of either
a~secular period change or apsidal motion. However,
it turned out that ---probably due to a~generally lower resolution of
older, photographic spectra, and a relatively small RV amplitude of
component Aa1--- there is a~significant difference in the semi-amplitude
between the older and more recent electronic spectra. Therefore, we derived
two other orbital solutions, for the older and more recent data, keeping
the orbital period fixed at the value from the solution for all data.
These are also listed in Table~\ref{solaa} and the final local systemic
velocities in Table~\ref{gam-aa1} are based on these two separate solutions.
The corresponding RV curves are shown in Fig.~\ref{rvc-aa1}.

\begin{figure}
\centering
\resizebox{\hsize}{!}{\includegraphics{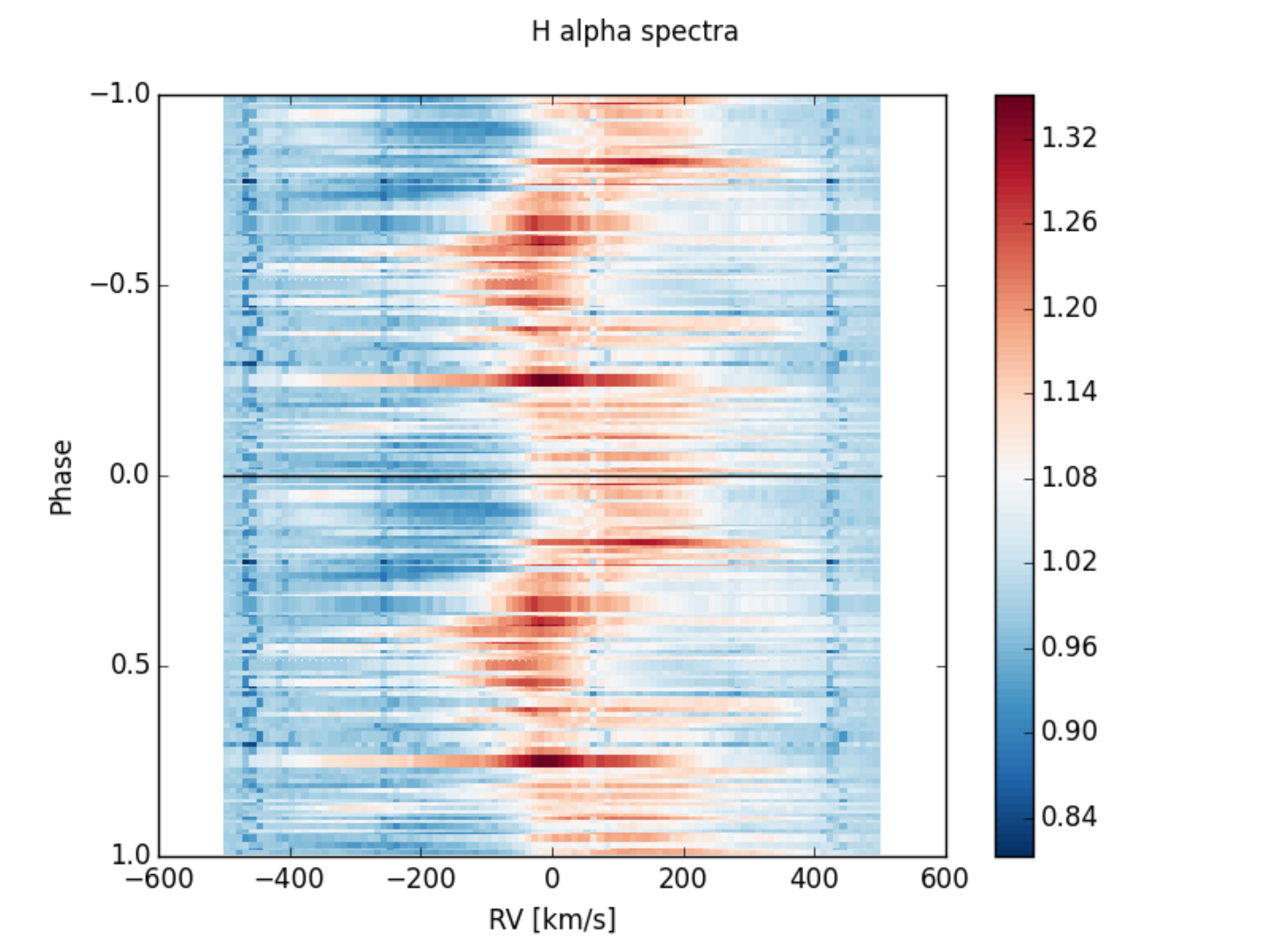}}
\resizebox{\hsize}{!}{\includegraphics{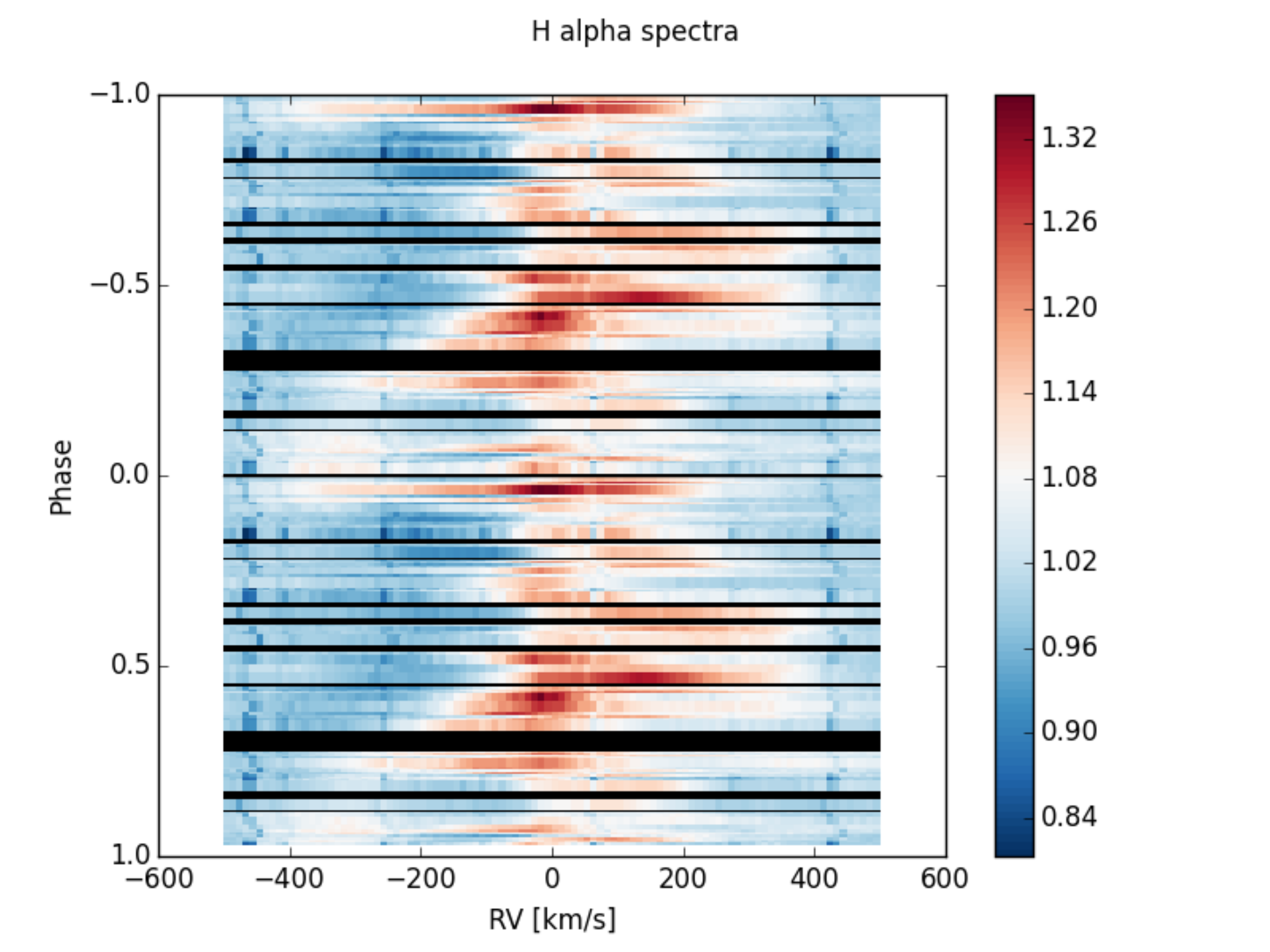}}
\caption{Trailed spectra of the \ha line profiles for all FEROS and CTIO
spectra plotted versus phase of the 20\fd7 orbit with phase zero at RV maximum of the component Aa1
from solution~C of Table~\ref{solaa} (top panel), and versus the phase of the
5\fd999 period for ephemeris (\ref{efphoI}) (bottom).}
\label{h3dyn}
\end{figure}

To investigate the character of relatively strong and variable
\ha emission, we also produced trailed spectra for all FEROS and
CTIO \ha line profiles versus the phases of both the 20\fd7 and 5\fd999 periods.
These are shown in Fig.~\ref{h3dyn} where we can see that the bulk of the emission
moves in orbit with component Aa1, although it is possible that a part of it
originates in other places within the system, probably being associated
with the circumstellar matter related to mass transfer from component Ac1
to Ac2. No trace of the secondary Aa2 is seen in the upper panel of the plot.

\begin{figure}
\centering
\resizebox{\hsize}{!}{\includegraphics{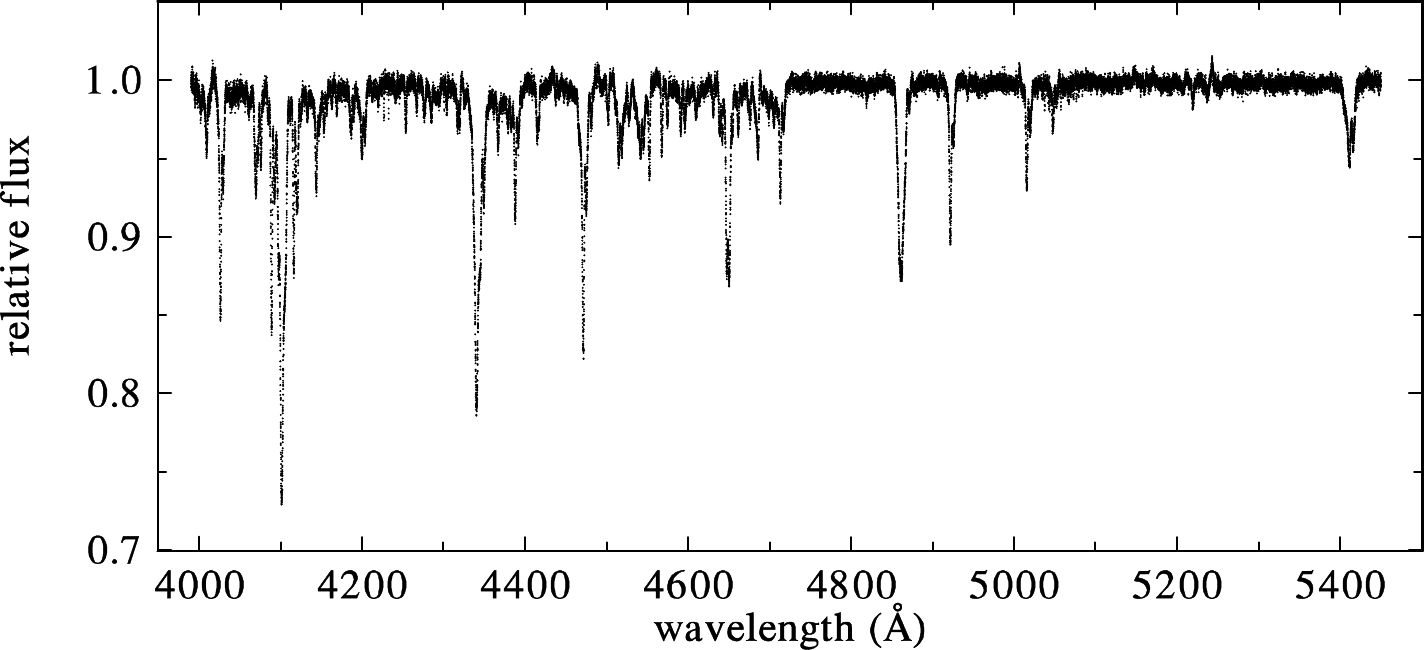}}
\caption{Blue part of the FEROS spectrum taken on RJD~54603.5911, with
lines of components Ac1 and Ac2 near one elongation.}
\label{feros2}
\end{figure}

\begin{figure}
\centering
\resizebox{\hsize}{!}{\includegraphics{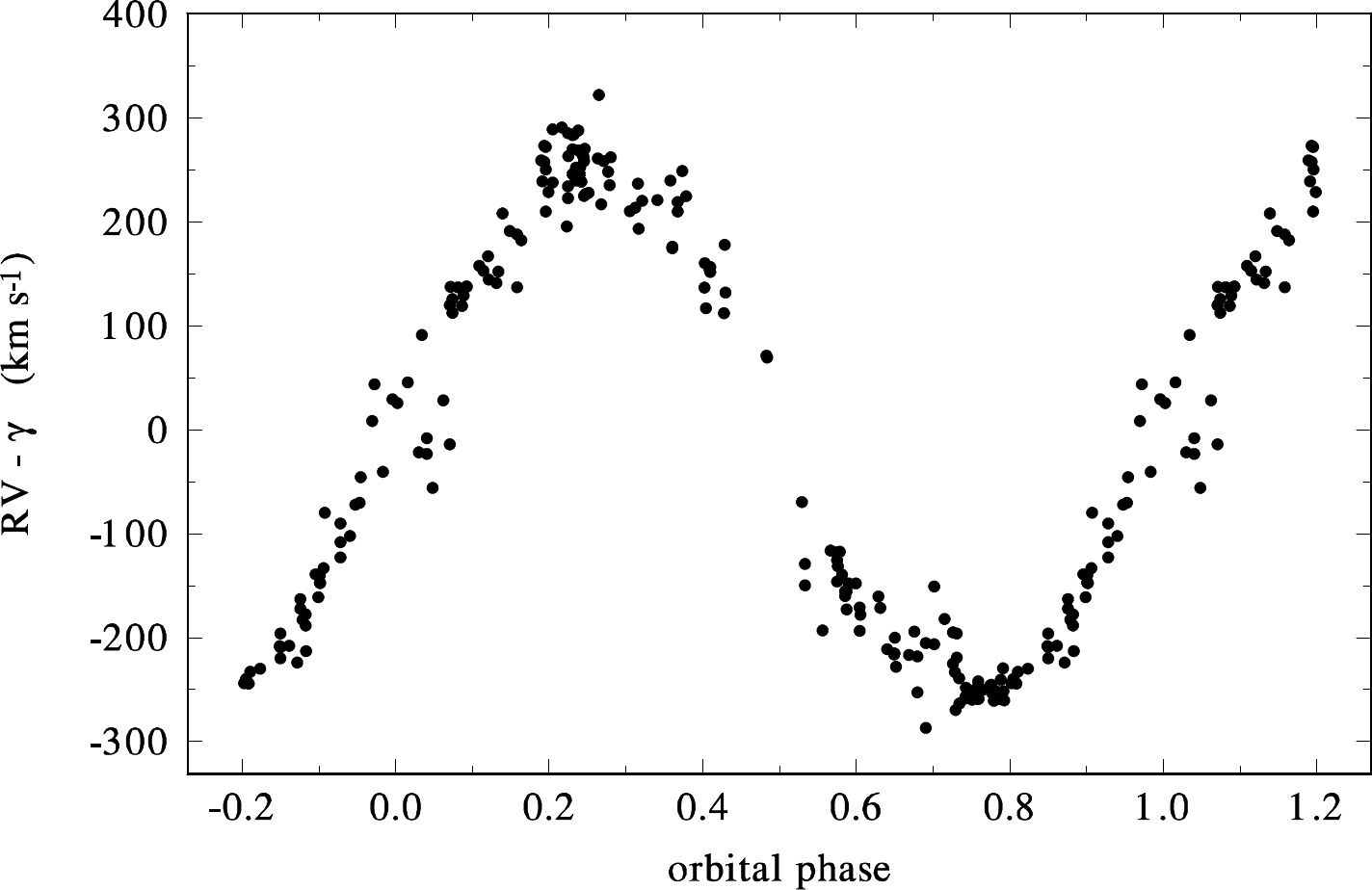}}
\caption{Radial-velocity curve of component Ac1 plotted with the ephemeris
of Table~\ref{solac}. We note that component Ac1 is eclipsed in
the secondary minimum (phase 0.5).}
\label{rvc-ac1}
\end{figure}

\begin{figure}
\centering
\resizebox{\hsize}{!}{\includegraphics{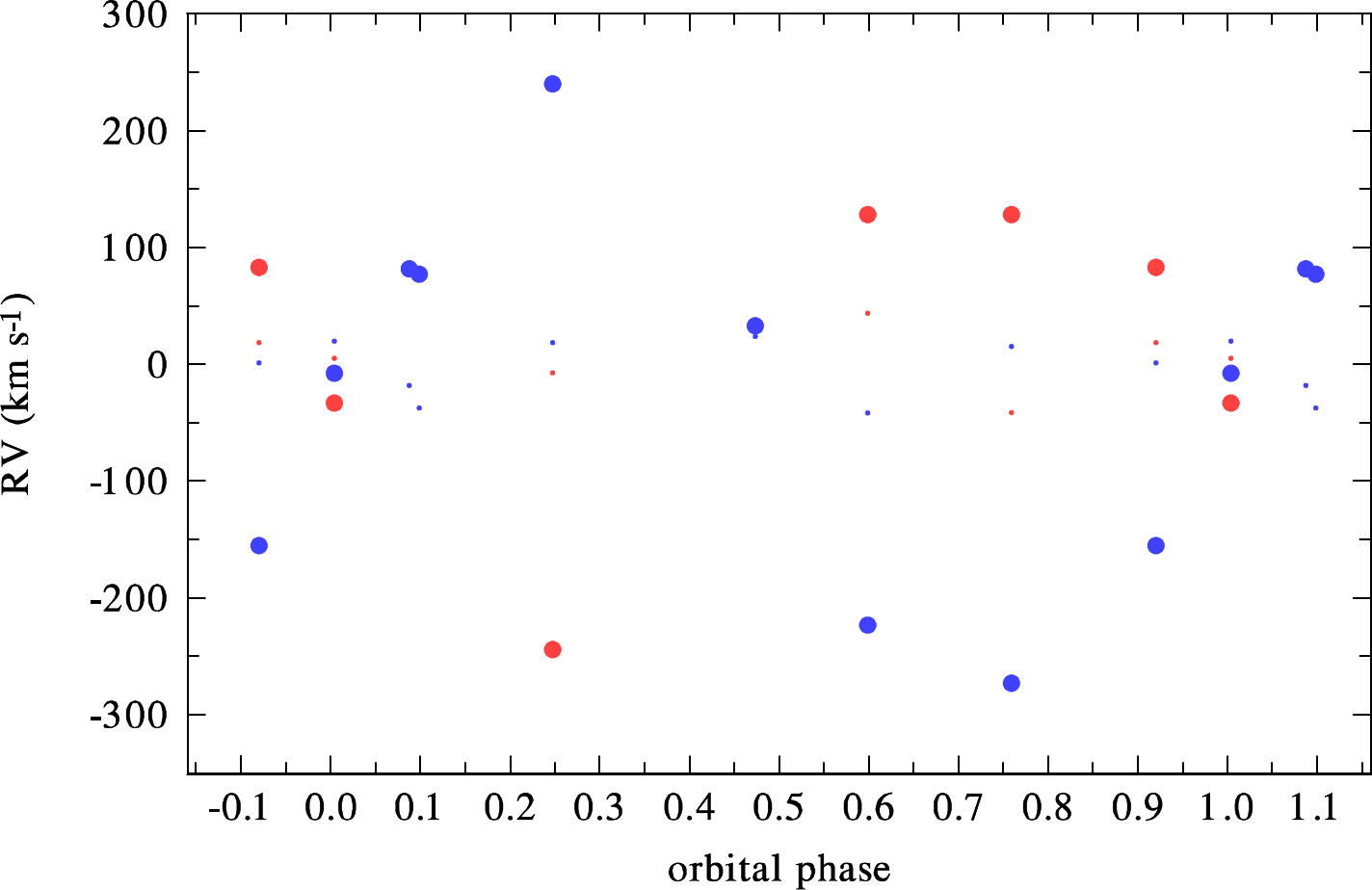}}
\caption{Radial-velocity curve of components Ac1 (blue) and Ac2 (red)
based on the \pyt fit of the blue FEROS spectra. Residuals are shown
by small dots.}
\label{rvpyt}
\end{figure}

\subsection{Orbit of the eclipsing components Ac1 and Ac2}
\subsubsection{Orbital solution, mass ratio, and radiative properties}
We first used all compiled and new \spefo RVs of component
Ac1 and it was possible to use the mean RV of all three blue lines,
\ion{He}{i}~4922 and 5016~\AA, and \ion{He}{ii}~5411~\AA. We also verified
that the RV curves derived from older and more recent data have the same 
(large) semi-amplitude; therefore a~single orbital solution for the
determination of local systemic velocities was used.
This solution is provided in Table~\ref{solac} and the local systemic velocities
with their rms errors and the range of the spectra are summarised
in Table~\ref{gam-ac1}. It is encouraging that this solution led
to a mean orbital period, which is consistent within the derived errors
with that from ephemeris~(\ref{efphoI}), obtained from fully independent 
photometric data. Values of the epoch of the primary minimum from
both solutions differ slightly more but this is not surprising given the presence
of the light-time effect and the different time distribution of photometric
and RV data.

 The less numerous (and much less accurate) RVs of component Ac2 cannot
be directly used for local $\gamma$ determination. However, we note that
there are two useful series of spectra, secured within relatively
short time intervals, for which we derived solutions with the orbital
period fixed and allowing for the mass ratio determination. The
resulting mass ratios $M_{\rm Ac2}/M_{\rm Ac1}$ were

\medskip
\centerline{FEROS+BESO over RJD~53739--54957: $1.29\pm0.18$}

\smallskip
and

\smallskip
\centerline{CTIO over RJD~58868--58908: $1.29\pm0.16$.}

\medskip
\noindent This provides the best estimate of the mass ratio of the
eclipsing pair  so far and confirms that the brighter, mass-losing component Ac1
is the less massive of the two, contrary to what \citet{stick} found from
the IUE spectra.

\begin{table}
\begin{center}
\caption{New circular orbital solution for all RVs of component Ac1 with
locally derived systemic ($\gamma$) velocities.
}
\label{solac}
\begin{tabular}{lrrrrrrrrccc}
\hline\hline\noalign{\smallskip}
Element            & Value        \\
\noalign{\smallskip}\hline\noalign{\smallskip}
$P$ (d)               &5.998693(12)  \\
$T_{\rm min\,I}$ (RJD)&49425.036(15) \\
$K_1$ (\ks)           & 253.2(2.5)   \\
No. of RVs            & 198          \\
rms (\ks)             & 26.61        \\
\hline\noalign{\smallskip}
\end{tabular}
\end{center}
\tablefoot{In the row `rms' we provide
the rms of 1~observation.}
\end{table}

We carried out yet another test using eight superb FEROS spectra in the blue
spectral region from 3990 to 5440~\AA\ (see one such spectrum
in Fig.~\ref{feros2}). We used the program \pyt
\citep[see][for details]{jn2016} to estimate the radiative properties,
relative luminosities, and also RVs of components Ac1, Ac2, and Aa1.
We note that \citet{sanch2017}, using different modelling approaches,
obtained relatively precise estimates of the flux brightness ratio of the two
close binaries $F_{\rm Ac}/F_{\rm Aa}$ (see their Table~3). These imply
the following estimates of the relative luminosity $L_3$ of component Aa:
$0.585\pm0.007$ for the Markov Chain Monte Carlo (MCMC) fit, $0.60\pm0.04$
for the LitPro fit, $0.546\pm0.001$ for the CANDID fit \citep{gall2015},
and $0.57\pm0.04$ for the average fit. \citet{sanch2017} adopted the average 
of the best fits by individual methods.
We are not sure that this is the best approach, considering the different levels
of sophistication of these models. In our opinion, the CANDID model should be adopted, which
also gives the lowest normalised $\chi^2$.
During the calculations, we omitted short segments around H$\gamma$ and
H$\beta$ lines, and the interval between 5130 and 5350~\AA,
affected by small emission features. Via artificial rectification, we also
removed stronger diffuse interstellar bands near 4428, 4726, 4762, and
4780~\AA. Uncertainties of the fit were estimated
with the MCMC modelling. The results are in
Table~\ref{pyter}. The radial-velocity curves of components Ac1 and Ac2
derived by \pyt are shown in Fig.~\ref{rvpyt}. The corresponding orbital
solution gives a mass ratio $M_{\rm Ac2}/M_{\rm Ac1}=1.25\pm0.14$ and
$K_{\rm Ac1}=255\pm15$~\ks, in good agreement with the results based
on all spectra.

\begin{table}[h!]
\begin{center}
\caption{Approximate radiative parameters of components Ac1, Ac2, and Aa1
estimated with the program \pyt from the 4990 -- 4400~\AA\ FEROS spectra.}
\label{pyter}
\begin{tabular}{cccccccc}
\hline\hline\noalign{\smallskip}
Quantity           &       Ac1    &       Ac2       &       Aa1  \\
\noalign{\smallskip}\hline\noalign{\smallskip}
\tef (K)           & 31950(351)   & 34310(982)      & 28940(355) \\
Best:              & 32050        & 33360           & 28760      \\
\lgg [cgs]         & 3.512(56)    & 3.95(10)        & 3.327(93)  \\
Best:              & 3.589        & 3.88            & 3.253      \\
$L$                & 0.281(21)    & 0.162(41)       & 0.556(52)  \\
Best:              & 0.273        & 0.140           & 0.585      \\
\vsin (\ks)        & 126(14)      & 327(31)         & 99(5)      \\
Best:              & 114          & 304             & 102        \\
\hline\noalign{\smallskip}
\end{tabular}
\end{center}
\end{table}

\subsubsection{Combined light-curve and RV-curve solution for system Ac}
We derived a final \phoebe solution using all calibrated \ubv\ data sets,
one TESS and one well-covered BRITE light curve, and adopting
the effective temperatures of 31980~K for component Ac1, and 34300~K
for component Ac2 from the \pyt solution (see Table~\ref{pyter}). During
each iteration, we modified the synchronicity parameter of
component Ac2 (the ratio between orbital period and the period of rotation
estimated from the instantaneous values of the radius, inclination and
\vsin derived with \pyte), the final value being about 3.9\,.
The corresponding solution is shown in Table~\ref{phoebe}. The system is
semi-detached, and the component Ac1 is filling the Roche lobe and is obviously
sending material towards Ac2.

\begin{table}
\centering
\caption{Combined radial-velocity curve and light-curve solution
with \phoebee.}\label{phoebe}
\begin{tabular}{lrlrl}
\hline\hline
Element& \multicolumn{4}{c}{Orbital properties (Ac)}\\
\hline
$P$\,(d)& \multicolumn{4}{c}{5.998682 fixed}\\
$T_{\rm Min\,I}$\,(RJD)& \multicolumn{4}{c}{$49425.0802\pm0.0023$}\\
$M_1/M_2$\,& \multicolumn{4}{c}{1.29 fixed}\\
$i$\,$(^\circ)$& \multicolumn{4}{c}{$87.8\pm0.8$}\\
$a$\,(\Rnom)$^{*}$& \multicolumn{4}{c}{$53.53\pm0.64$}\\
$L^{\rm Aa}_{\rm V}$      &\multicolumn{4}{c}{$0.522\pm0.001$}\\
$L^{\rm Aa}_{\rm B}$      &\multicolumn{4}{c}{$0.470\pm0.001$}\\
$L^{\rm Aa}_{\rm U}$      &\multicolumn{4}{c}{$0.403\pm0.003$}\\
$L^{\rm Aa}_{\rm TESS}$   &\multicolumn{4}{c}{$0.503\pm0.001$}\\
$L^{\rm Aa}_{\rm BRITE}$  &\multicolumn{4}{c}{$0.521\pm0.001$}\\
\noalign{\smallskip}\hline
& \multicolumn{4}{c}{Component properties}\\
& \multicolumn{2}{c}{Component Ac1}& \multicolumn{2}{c}{Component Ac2}\\
\hline
\tef\,(K)& \multicolumn{2}{c}{31980 fixed}& \multicolumn{2}{c}{34300 fixed}\\
$\log g_{\rm}$ [cgs]& 3.27 & &3.94\\
$M$\,(\Mnom)$^{*}$& 25.0& & 32.2\\
$R$\,(\Rnom)$^{*}$& 19.1& &10.0\\
$M_{\rm bol}$   &$-9.11$& &$-8.00$ \\
$L_{\rm V}$      &0.366 &&0.112 \\
$L_{\rm B}$      &0.404 &&0.126 \\
$L_{\rm U}$      &0.454 &&0.143 \\
$L_{\rm TESS}$   &0.374 &&0.123 \\
$L_{\rm BRITE}$  &0.369 &&0.110 \\
\hline
\end{tabular}
\tablefoot{
$^*$ Masses and radii are expressed in nominal solar units, see
\citet{prsa2016}.
}
\end{table}

\begin{figure}
\centering
\includegraphics[width=0.9\columnwidth]{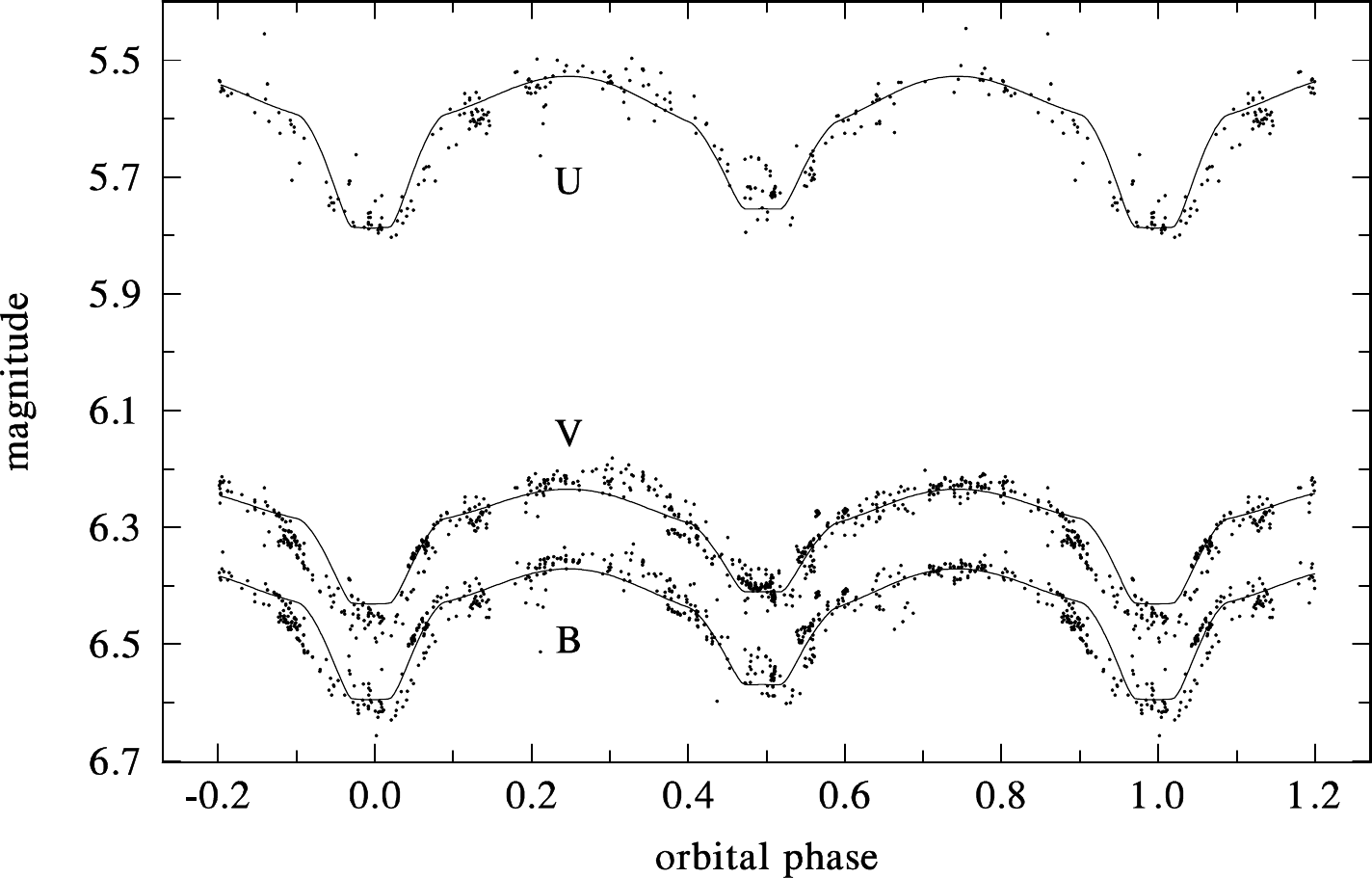}
\includegraphics[width=0.9\columnwidth]{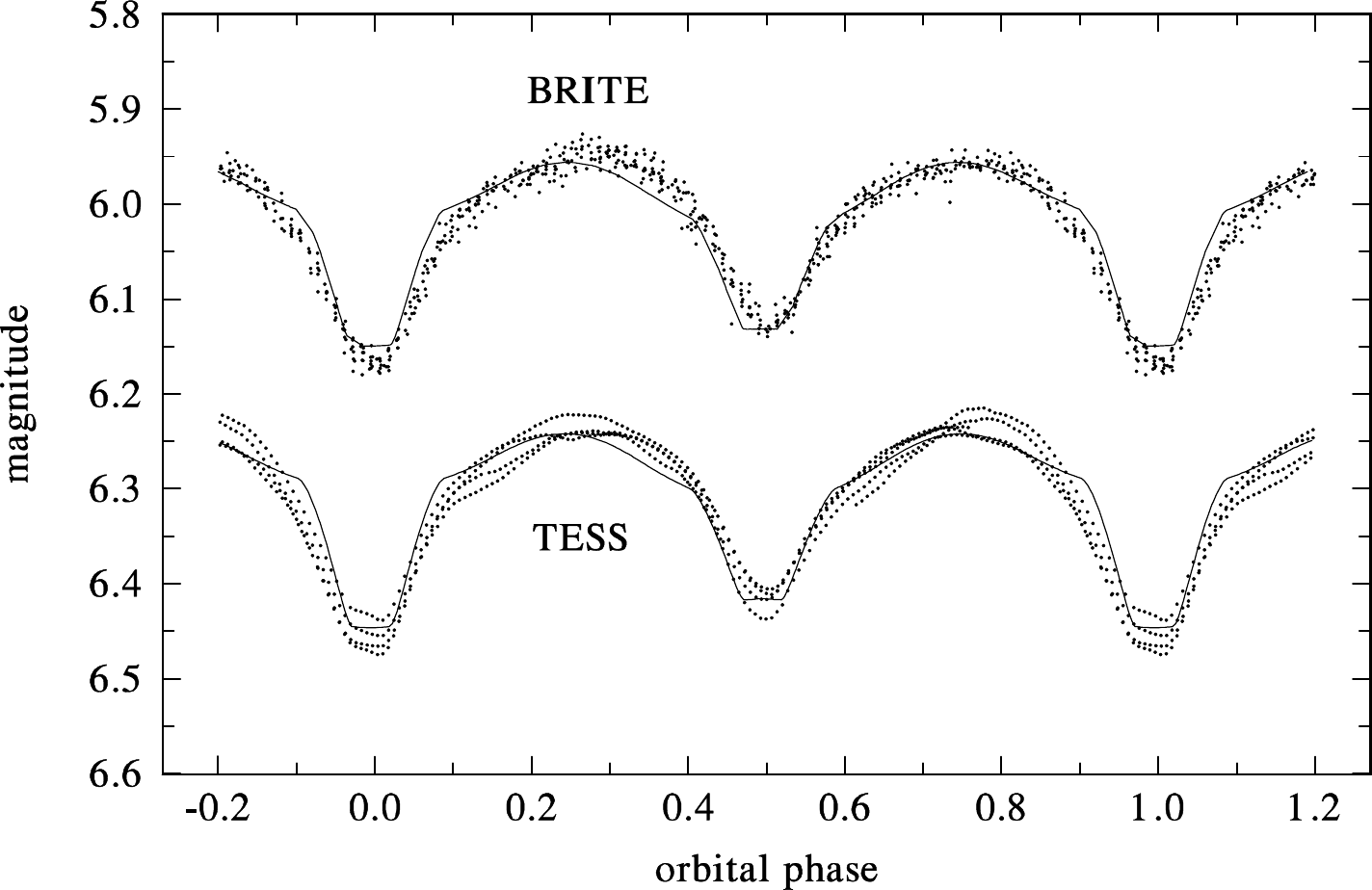}
\caption{Fit of \ubv\ (top) and BRITE and TESS light curves
(bottom) with \phoebe1. Phases are calculated with ephemeris~(\ref{efphoI}).}
\label{screen}
\end{figure}

The relative contribution of the third light
(i.e. flux from the subsystem Aa, which is predominantly from Aa1) is decreasing
towards shorter wavelengths, confirming that the effective temperature
of component Aa1 is lower than those of components Ac1 and Ac2.
There is relatively~satisfactory agreement between the results from \pyt
and our \phoebe solution. The fit of the light curves used is shown
in Fig.~\ref{screen}.

The combined solution of the calibrated light curves and a RV curve of
component Ac1 for the fixed mass ratio 1.29 presented in Table~\ref{phoebe}
provides the following standard \ubv\ values for individual
components (if one uses the \ubv\ values of the whole system in elongations
defined by the model fit):

\smallskip
\centerline{Ac1: $V=7$\m325, \bv$=0$\m029, \ub$=-0$\m969,}
\centerline{Ac2: $V=8$\m615, \bv$=0$\m008, \ub$=-0$\m983,}
\centerline{Aa1: $V=6$\m940, \bv$=0$\m251, \ub$=-0$\m676.}

\noindent A standard dereddening for the components of the eclipsing
subsystem Ac then gives $E$(\bv) of 0\m357 and 0\m337 for components
Ac1 and Ac2, respectively. Using the mean of these two estimates,
$E$(\bv)$=0$\m347, one obtains

\smallskip
\centerline{Ac1: $V_0=6$\m17, (\bv)$_0=-0$\m32, (\ub)$_0=-1$\m22,}
\centerline{Ac2: $V_0=7$\m53, (\bv)$_0=-0$\m34, (\ub)$_0=-1$\m24.}

\noindent Together with the adopted effective temperatures and bolometric
magnitudes and using \citet{flower96} bolometric corrections, one obtains
$M_{\rm V}=-6$\m11 and $-4$\m85 for Ac1 and Ac2, respectively, which implies
a distance towards the system of about 2860-3000~pc, in broad agreement
with the distance of the Collinder~228 cluster.

If component Aa1 is a genuine supergiant, then for its estimated
\tef = 28900~K, it should have a deredenned (\bv)$_0=-0$\m25 according
to calibration by \citet{flower96}. For the above-quoted deduced
\bv$=0$\m251, this would imply a significantly higher $E$(\bv) of
about 0\m50.

We note that the parallax of \qz as given in Gaia DR2 and EDR3 (without
account of the multiplicity of the object) seems to be overestimated.
The surrounding close-by stars from the cluster have parallaxes that are 
about~half those of \qz itself. The distance
to the cluster was estimated from 29 O stars by \citet{shull2019}
to be of $2.87\pm0.73$~kpc from Gaia DR2, and $2.60\pm0.28$~kpc from
photometry, which are similar to what we estimate from the combined solution above.
However, it must be kept in mind that nebular emission is known for the
Carina nebula and that the observed reddening could be partly affected
by that or even by the circumstellar matter within QZ~Car itself. Therefore,
it may not provide a good estimate of the distance to the system.

The radius of component Ac1 and the orbital inclination from the combined
solution presented in Table~\ref{phoebe} would imply \vsin = 161~\kms for this component,
while \pyt gives $126\pm14$~\ks. The combined solution also predicts
\lgg = 3.27 [cgs] for component Ac1, while the \pyt value is $3.51\pm0.06$.
The agreement in this case is not ideal.

\subsection{Wide orbit}
\begin{figure}
\centering
\resizebox{\hsize}{!}{\includegraphics{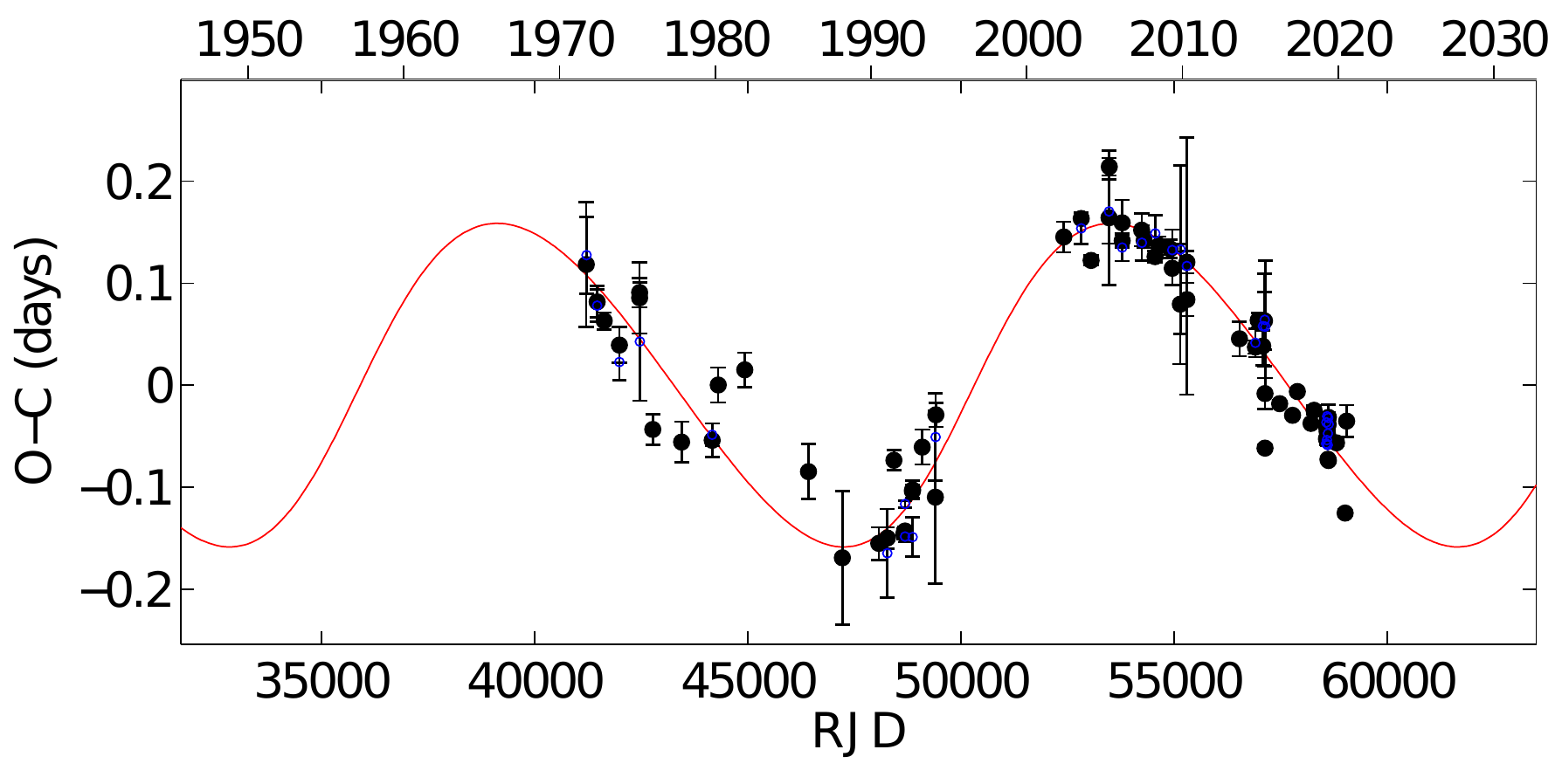}}
\resizebox{\hsize}{!}{\includegraphics{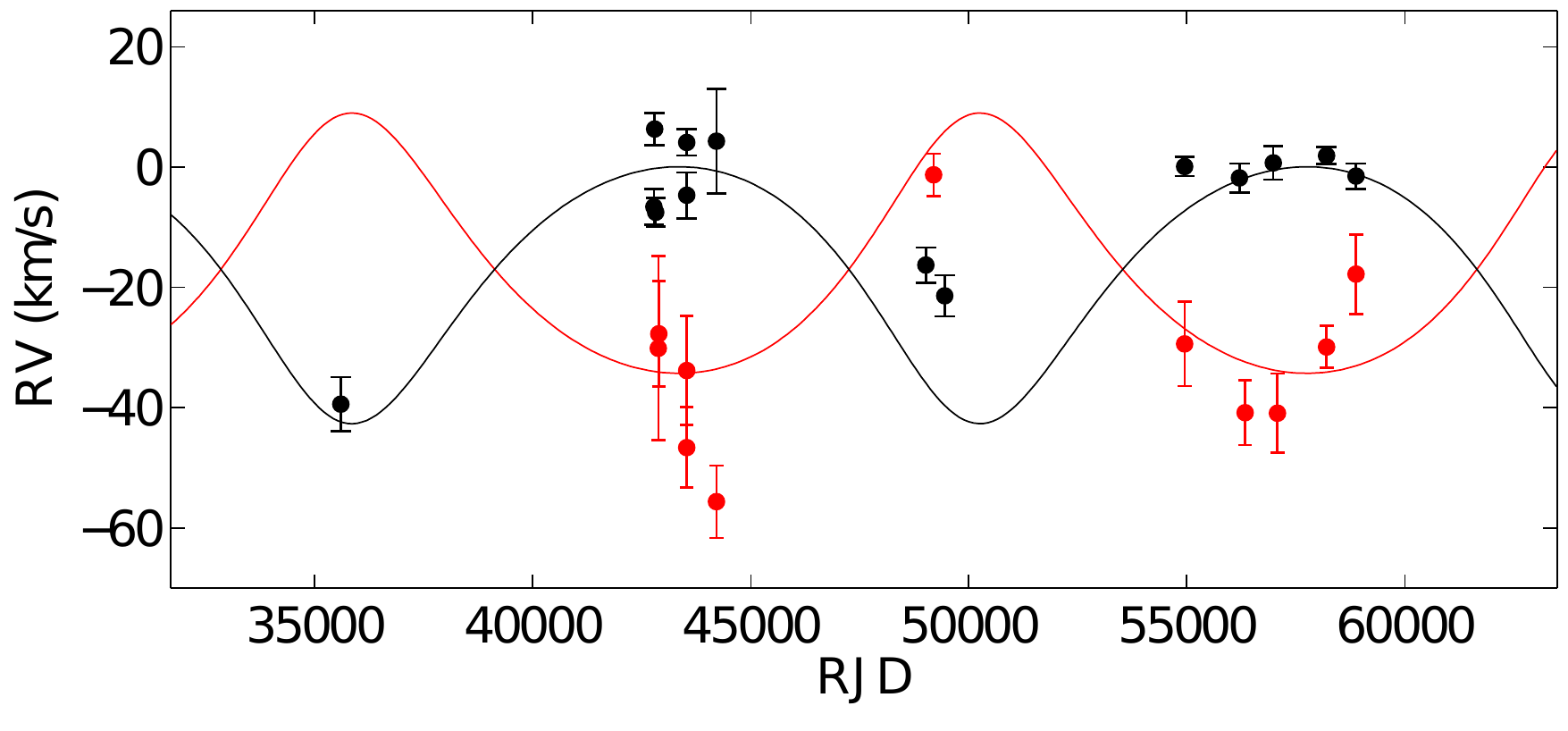}}
\resizebox{\hsize}{!}{\includegraphics{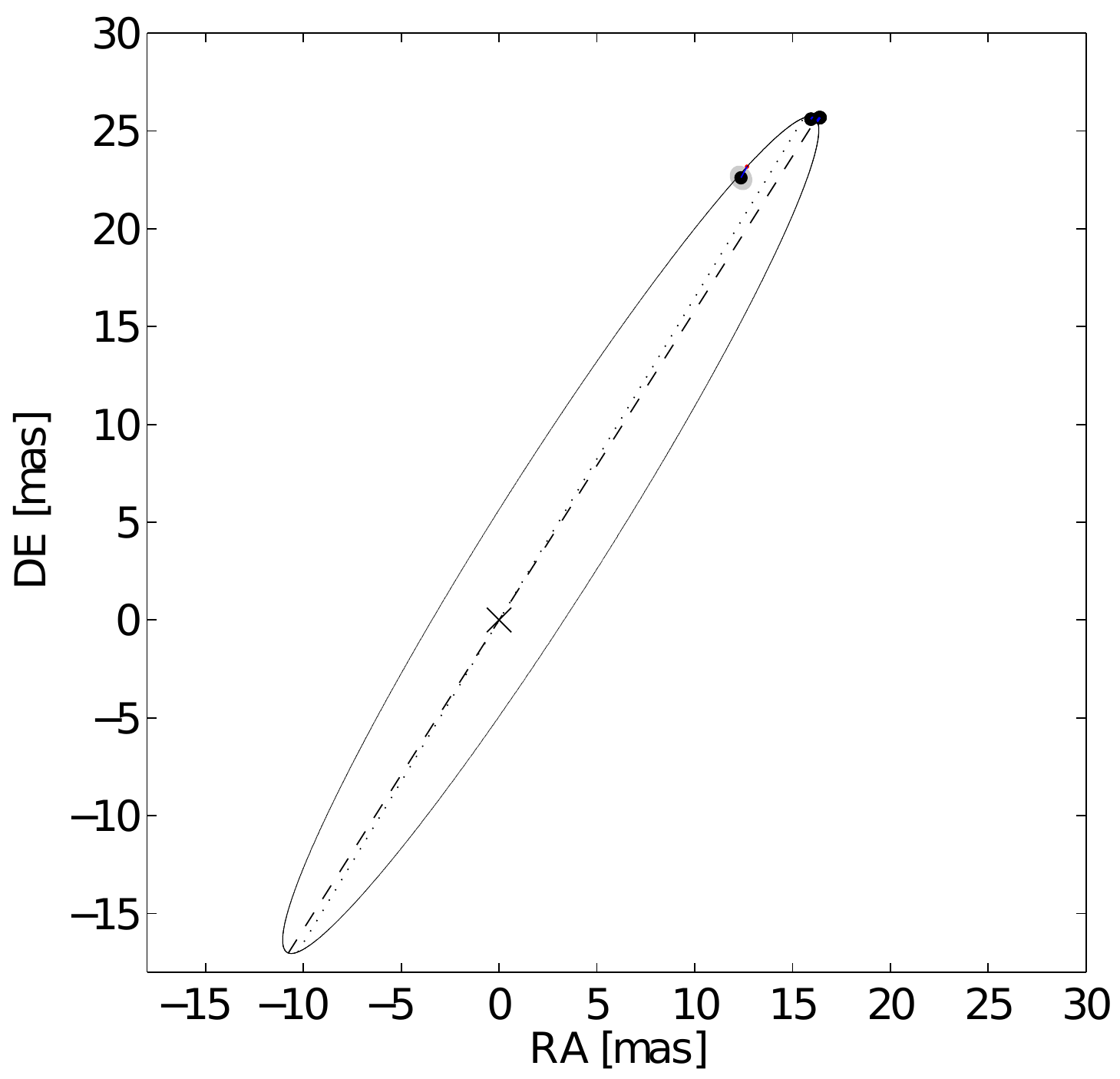}}
\caption{A combined fit of data for the long period of 14403~d.
Top panel: The \oc\ deviations from the times of primary minima.
Middle panel: Systemic velocities of components Aa1 (black) and Ac1 (red).
Bottom panel: The corresponding spatial long orbit,
which fits the four precise interferometric observations, with the pairs
Aa and Ac resolved.}
\label{longorb}
\end{figure}

The current estimates of the mutual orbital period of the two pairs, Aa and Ac,
are still rather uncertain, ranging from about 32 to 50 years
($11700-18300$~d). There were several attempts to estimate the orbital
elements of the long orbit, mainly based on the eclipse timing variation
(ETV) analysis \citep{MLDA, walker2017,black2020}. However, as only one
minimum and one maximum of the \oc\ diagram have so far been covered
(see the top panels of Fig.~\ref{longorb}), there is a rather large range
of possible values of the period of the mutual orbit of pairs Aa1-Aa2
and Ac1-Ac2.

Having the richest and the most homogeneous data sets, we attempted
to obtain a more accurate estimate of the elements of the wide orbit.
To demonstrate the inherent uncertainties
in the determination of the times of binary eclipses, we derived them in
two different ways:

  (1) We derived local estimates of the times of the primary eclipse of
the Ac pair using a formal light-curve solution with \phoebe1 over limited
time intervals and keeping the orbital period fixed at the value
from the complete solution; see ephemeris (\ref{efphoI}). This may provide
a more robust estimate of the minima than the estimates for the individual,
often not ideally covered minima. The results are provided in
Table~\ref{minima}. We then derived the eclipse timing residuals (ETR) with
respect to ephemeris~(\ref{efphoI}).

\begin{table*}[!ht]
\begin{center}
\caption{Formal `orbital' solutions for the mutual orbit of components Aa
and Ac.}\label{solong}
\begin{tabular}{lrrrrrrrrccc}
\hline\hline\noalign{\smallskip}
Element                 & Sol. 1    & Sol. 2   & Sol. 3   & Sol. 4      & Sol. 5  \\
\noalign{\smallskip}\hline\noalign{\smallskip}
$P$ (d)                 &14234(419) &14624(149)&14403(178)&15011(1926)  &12008(245)  \\
$T_{\rm periast.}$ (RJD)&52002(1288)&49564(430)&50072(547)&51501(4788)  &55774(1173) \\
$\omega$ ($^\circ$)     &312(28)    &251.6(9.9)&265(12)   &195(115)     &332(45)     \\
eccentricity            &0.189(98)  &0.188(35) &0.204(40) &0.19 (fixed) &0.28(12)    \\
Semi-amplitude (\ks/d)  &0.139(12)  &0.1684(48)&0.1586(70)&16(11)       &25.9(3.2)   \\
Mean value (\ks/d)      &0.023(87)  &0.015(5)  &0.022(5)  & $-12.3(3.2)$&$-15.2(3.8)$\\
No. of obs.             & 32        & 64       & 96       & 14          &11          \\
rms (\ks)               &0.036      &0.021     &0.028     & 5.30        &4.36        \\
\hline\noalign{\smallskip}
\end{tabular}
\end{center}
\tablefoot{Solution~1 is for the ETRs based on the primary minima derived
from locally fitted complete light curves, solution~2 on ETRs derived
from the individual primary and secondary minima, solution (3) is a combined
solution for the instants of minima derived by both methods, solution~4 is
based on the locally derived systemic ($\gamma$) velocities of component Aa1,
and solution~5 on the locally derived $\gamma$ velocities of component Ac1.
Eccentricity $e$ had to be fixed for solution~4 as it turned out
to be  unconstrained in a free solution.}
\end{table*}

 (2) Alternatively, the individual light curves were also analysed in
the following, perhaps more customary way. Individual photometric datasets
were used separately as input for the~semi-automatic routine 
automatic fitting procedure \citep[AFP; see][]{zasche2014}.
This routine returns individual times of the primary and secondary
eclipses using a light-curve template provided by {\tt PHOEBE}. It therefore
provides up to twice as many eclipse times. However, their accuracy
depends on how well the individual minima are covered
by observational data and it can often be slightly poorer than the accuracy of the
minima defined by the whole local light curves. For that reason,
only reasonably well covered minima were used. This is why the corresponding
ETRs show a lower scatter than those derived using the first method.
The instants of the primary
minima derived in this way are shown in Table~\ref{pminima}, and those for the secondary
minima are shown in Table~\ref{sminima}.

 We attempted to estimate the period and some other elements of
the wide orbit for the ETRs derived using both of the above-mentioned ways.
We modelled individual quantities formally as a motion
in an eccentric orbit using \fotele. These are listed as solutions (1) and (2)
in Table~\ref{solong}. As both these solutions led to similar values
of the long orbital period and eccentricity, we combined both sets of
minima and derived a joint final solution, which is presented in Table~\ref{solong}
as solution~3.

We then derived orbital solutions for the wide orbit using the local
systemic ($\gamma$) velocities of both components, Aa1 and Ac1.
These are also listed in Table~\ref{solong} as solutions~4 and 5.
A~free convergence of all elements for the local $\gamma$ velocities of
component Aa1 led to an unconstrained value of orbital eccentricity of
$0.62\pm7.8$, and so we derived solution (3) for Aa1 with eccentricity fixed at
the value obtained from both ETRs fits.

Our results illustrate the remaining, non-negligible uncertainties of
the current knowledge of the elements of the long orbit but show that
the orbital period is probably close to about 14500~d. Very regrettably,
no RVs of components Aa1 and Ac1 are available for the time interval
from RJD~51000 to RJD~53000, where there should be the largest difference
between $\gamma$ velocities of systems Aa and Ac. The value of the orbital
amplitude of component Aa in the wide orbit thus strongly depends on the early
photographic spectra. From the results in Table~\ref{solong}, one can only
conclude that the masses of systems Aa and Ac are of the same order of
magnitude, system Aa being probably more massive than Ac.

There is also another, and potentially very important source of
information, namely the new interferometric observations obtained recently by
\citet{sana2014} and \citet{sanch2017}, who directly resolved the
two pairs. Considering the large uncertainty of the observation obtained with 
the Sparse Aperture Masking (SAM) instrument by \citet{sana2014},
we decided to skip this one and we used only the two other data points.
Moreover, we used two more recent interferometric observations 
(from March 14, 2017, and April 27, 2017) secured with the GRAVITY instrument 
attached to Very Large Telescope Interferometer (VLTI), which we found
in the ESO archive.\footnote{Their reduction is described
in Paper~II, where the resulting astrometric data are also used.}
These four positional measurements and our complete sets of the ETRs
discussed above were used to get a~combined solution for the long orbit.
As only four positional observations secured over a limited time-interval
are insufficient to define the wide orbit reliably, even more so given that all of them exhibit
a~non-negligible correlation between the separation and the position-angle
error bars, we derived a constrained solution adopting the period and
eccentricity from the light-time effect fit, that is solution~3 of
Table~\ref{solong}. Corresponding ETRs are compared to the light-time
effects and to systemic velocities in the top and middle panels of
Fig.~\ref{longorb}. The light-time effect due to the mutual orbit of the
two pairs is clearly visible, as is the antiphase variation of
the $\gamma$ velocities of components Aa1 and Ac1.
The projection of the orbit is shown in the bottom panel of Fig.~\ref{longorb},
where the error ellipses are also plotted (we note that for the second one,
the observation from GRAVITY \citep{sanch2017}, the ellipse is so small that
it cannot even be seen around the black dot).

The solution gives an inclination of the wide orbit
$i_3=96^\circ8\pm8^\circ1$, $\Omega_3=147^\circ6\pm6^\circ5$ and
requires a parallax of 0\farcs00045, implying a distance to the system
of over 2200~pc. The mass function of 14.17~\Mnom\  then implies the mass of
the whole Aa system of 57.1~\Mnom. Using the observed mass function
for the 20\fd7 spectroscopic Aa1-Aa2 orbit from solution~C of
Table~\ref{solaa} and assuming approximate coplanarity of all three orbits,
one obtains $M_{\rm Aa1}=47.6$~\Mnom\ and $M_{\rm Aa2}=9.5$~\Mnom.
Our nominal model thus leads to the total masses of systems Aa and Ac,
which are almost equal.

\section{Discussion of the emerging picture of the quadruple system}
  We made attempts to disentangle the eight good-quality blue FEROS
spectra using the program \korel \citep{korel1, korel2}. We were unable to
find a unique and stable solution, which is unsurprising given that those spectra
only cover a limited part of the mutual orbit of the Aa and Ac pairs.
However, there are indications that the emission components of
H$\beta$ and H$\gamma$ lines are associated with the mass-gaining
component Ac2. This in turn means that what we actually observe for Ac2
could be a pseudophotosphere, that is, inner optical parts of the accretion disk.

   We also tried to analyse the residuals from the light-curve solutions,
separately for the $V$ magnitude data, and TESS and BRITE photometries.
The only detected frequencies were close to the one-year alias of the orbital
period and its harmonics. We conclude that the obvious changes in the
shapes of the light curves from one cycle to another are stochastic by
nature, probably related to changes in the circumstellar matter.
Similar changes are known for $\beta$~Lyr \citep{larsson70,hec96}
and some other interacting binaries.

   There is also some indication of a slight secular change of the total
brightness of the object in the standardised yellow magnitude but
confirmation will require future accurate differential observations 
relative to a~non-variable comparison star. For completeness, we also mention
that the residuals from the fit of light-time changes based on local fits
of complete light curves (solution~2 of Table~\ref{solong}) can be reconciled
with a period of 651~d.

  Current knowledge of the physical properties of the Aa system is still
rather uncertain because we do not see the spectrum of its secondary Aa2.
Consequently, there is no RV curve of the companion and no direct estimate
of the binary masses. Also, the orbital inclination is not known. We only have
the  mass function and can estimate the system properties if we adopt some
value of mass for component Aa1.

\section{Conclusions}
We collected and critically evaluated a large body of observational data
and obtained much improved orbital solutions for subsystems Ac and Aa of the nine-component system HD~93206.
We estimated the mass ratio of the Ac system to be 1.29, confirming that
the more massive object is a mass-gaining and less luminous secondary,
which is eclipsed in the primary minimum. This system is
obviously in the still rather rapid phase of mass transfer between
the components, and is observed relatively shortly after the mass-ratio
reversal. It is also conceivable that even the Aa1 component of the 20\fd7
spectroscopic binary is actually a shell star, a product of previous mass
transfer in that system.

   Our analysis, in the context of what is currently known, shows that a better understanding of this unusual system
will require not only new series of high-S/N spectra covering also the
blue part of the optical spectrum but also more sophisticated models.
A first step in this direction is the accompanying Paper~II (Bro\v{z}
at al.).

\begin{acknowledgements}
We thank J.A.~Nemravov\'a, who left the astronomical
research in the meantime, for her contributions to this study and for the
permission to use her Python programs, \pyt and several auxiliary ones.
We also profitted from the use of public versions of the program
\phoebe~1 by A.~Pr\v{s}a and T.~Zwitter, and the programs \fotel and \korele,
written by P.~Hadrava.
We acknowledge the use of photometric observations secured by the late
Harry Williams at his private Milton Road Observatory and the FRAM $V$
photometry kindly provided by M.~Ma\v{s}ek.
The research of P.H., M.B., M.W., and J.J. is supported by
the grant GA19-01995S of the Czech Science Foundation.
B.B. is supported by the NSF grant AST-1812874.
J.L.-B. acklowledges support from FAPESP (grant 2017/23731-1).
H.B. acknowledges financial support from
the Croatian Science Foundation under the project 6212 ``Solar and
Stellar Variability".
This work has made use of data from the European Space Agency (ESA)
mission {\it Gaia} (\url{https://www.cosmos.esa.int/gaia}), processed by
the {\it Gaia} Data Processing and Analysis Consortium (DPAC,
\url{https://www.cosmos.esa.int/web/gaia/dpac/consortium}). Funding
for the DPAC has been provided by national institutions, in particular
the institutions participating in the {\it Gaia} Multilateral Agreement.
The paper includes data collected by the TESS mission, which are publicly
available from the Mikulski Archive for Space Telescopes (MAST). Funding for
the TESS mission is provided by NASA's Science Mission directorate.
We gratefully acknowledge the use of the electronic databases: SIMBAD at CDS,
Strasbourg and NASA/ADS, USA. We also acknowledge the use of the WDS catalogue
of visual binaries and multiple systems and thank to Brian Mason for providing
us with individual astrometric observations of the components of HD 93206
that are collected in the catalogue.
\end{acknowledgements}

\bibliographystyle{aa}
\bibliography{qzcar}

\begin{appendix}
\section{Published orbital solutions for the two subsystems}
\label{apt}

  The orbital solutions for the spectroscopic 20\fd7 Aa1-Aa2 binary are
summarised in Table~\ref{arvsol}, and those for the eclipsing 6\fd0 Ac1-Ac2
binary are shown in Table~\ref{crvsol}.

\begin{table*}
\begin{flushleft}
\caption{Published orbital solutions for the spectroscopic subsystem Aa1-Aa2.}
\label{arvsol}
\begin{tabular}{lrrrrrrrrccc}
\hline\hline\noalign{\smallskip}
 Solution:         & 1 \ \ \   & 2  \  \   & 3A  \ \ \  &3B \ \      &4 \ \   &5 \ \          &6 \ \ \\
\noalign{\smallskip}\hline\noalign{\smallskip}
$P$ (d)            &20.72(2)   &20.72 fixed&20.73(1)    &20.73 fixed &20.73596&20.7374(12)    &20.73596 fixed\\
$T_{\rm periastr.}$ (RJD)&42529.8    &    --     &42530.0(0.7)$^\ast$&42530.4(1.5)&42530.49&42530.17(0.47) &42530.49 fixed\\
$T_{\rm inf.conj.}$ (RJD)&   --      &43532.72   &    --      &    --      &  --    &   --          & --           \\
 $e$               &0.34(4)    &0.34 fixed &0.34(6)     &0.26(8)     &0.342   &0.398(53)      &0.42(0.1)     \\
$\omega$ ($^\circ$)&126(11)    &  126 fixed&131(15)     &134(16)     & 143.6  &131.9(9.8)     &142(9)        \\
$K_1$ (\ks)        & 48(2)     &   48 fixed&49(4)       &47(4)       & 49.6   &46.4(2.9)      &53.6(3.2)     \\
$\gamma$ (\ks)     &$-8$(2)    &$+$4       &$-$7(2)     &$+$7(3)     &$-$19.1 &$-$2.8(1.9)    &$+$0.8(2.3)   \\
No. of RVs         & 29        &  16       &27          &28          & 25     &9              &34            \\
rms (\ks)          &  --       &  --       &10          &13          & 6.9    &12.7           & 9            \\
\hline\noalign{\smallskip}
\end{tabular}
\end{flushleft}
\tablefoot{References coded in the row `Solution':
1... \citet{MC79};
2... \citet{LMS};
3A... \citet{MC}, \ion{He}{i} lines;
3B... \citet{MC}, \ion{Si}{iv} and \ion{N}{iii} lines;
4... \citet{MLDA};
5... \citet{stick};
6... \citet{walker2017}.\\
$^\ast$) Their value 42520.0 is an obvious misprint.
}
\end{table*}

\begin{table*}
\begin{flushleft}
\caption{Published orbital solutions for the eclipsing subsystem Ac1-Ac2.}
\label{crvsol}
\begin{tabular}{lrrrrrrrrccc}
\hline\hline\noalign{\smallskip}
 Solution:         & 1 \ \ \  & 2  \ \       & 3A  \ \    &3B \ \      &4 \ \     &5 \ \         &6 \ \ \\
\noalign{\smallskip}\hline\noalign{\smallskip}
$P$ (d)            &5.9965(15)&5.9983(9)     &5.9980(9)   &5.9976(9)   &5.9991    &5.99843(10)   &5.9991 fixed\\
$T_{\rm periastr.}$ (RJD)&   --     &    --        &42108.7(0.3)&42108.7(0.5)&  --      &43512.25(0.21)&  --          \\
$T_{\rm min.I}     $ (RJD)&43235.9  &42472.64(0.10)&     --     &            &48687.16  &   --         &48687.16 fixed\\
 $e$               &0.04\p0.04& 0.0 assumed  &0.09\p0.05  &0.26\p0.12  &  0.0     &0.10(2)       &    0.0       \\
$\omega$ ($^\circ$)& 5\p48    &   --         &  35\p20    & 20\p18     &   --     &24.7(12.7)    &   --         \\
$K_1$ (\ks)        & 256\p10  &  260\p10     & 255\p6     & 284\p27    &259.7(2.0)&249.5(5.1)    &255           \\
$K_2$ (\ks)        &    --    &   --         &    --      & --         &   --     &266.6(8.7)    &   --         \\
$\gamma$ (\ks)     &$-26$\p7  &$-36$\p11     &$-$34\p8    &$-$47\p18   &$+$1.7    &$-$42.8(4.3)  &$-$46         \\
No. of RVs         & unknown  &  16          &   17       &14          & 11       &   9          &  19          \\
rms (\ks)          &not given &  20          &   35       &34          &  ?       & 24.6         &  11          \\
\hline\noalign{\smallskip}
\end{tabular}
\end{flushleft}
\tablefoot{References coded in the row `Solution':
1... \citet{MC79};
2... \citet{LMS};
3A... \citet{MC}, \ion{He}{i} lines;
3B... \citet{MC}, \ion{Si}{iv} and \ion{N}{iii} lines;
4... \citet{MLDA};
5... \citet{stick};
6... \citet{walker2017}.
}
\end{table*}

\section{Details of the spectral data sets, reduction and measurements, and
comments on archival RVs}\label{apa}
Initial reductions of FEROS and BESO spectra were carried out by A.N., and those
of CTIO CHIRON spectra were carried out by R.C.-H. and W.F.
Normalisation, removal of residual cosmics, and flaws and
RV measurements of all spectra were carried out within the program \spefo
\citep{sef0,spefo}, namely the latest version 2.63 developed
by J.~Krpata \citep{spefo3}. \spefo displays direct and flipped traces of
the line profiles superimposed on the computer screen that the user can slide
to achieve a precise overlapping of the parts of the profile to be measured.
 Following our usual approach, we also measured a~selection of good telluric
lines in the red spectral regions to obtain an additional fine correction
of the RV zero-point of each spectrogram. The wavelength zero-point of BESO
spectra was also checked via \ion{Na}{i}~5889 and 5895~\AA\ interstellar lines.

Here we provide comments on the RV data sets compiled from
the astronomical literature, which are listed in Table~\ref{jourv}.
File A: These RVs are from the Radcliffe 2-prism photographic spectra secured
in 1956, calibrated to the IAU RV standards. All RVs of \qz are denoted as
uncertain by the authors. Two of the four spectra were measured repeatedly by different
measurers, revealing RV differences of up to 14 \ks. It is not a~priori clear to
which component they refer. They might be reconciled
roughly with the 20\fd734 orbit of component Aa1, but with a large scatter and
very negative mean RV. Plotting them versus phase of the 5\fd9987 period
resulted in a scatter diagram.
The \ion{Ca}{ii} interstellar lines were also measured, their mean RV
being $+0.4\pm1.6$~\kms from six measurements.

File B: These three photographic RVs have a small RV range from $-11$ to
0~\kms and do not follow either the 20\fd735 or 5\fd9987 orbit, as
already demonstrated by \citet{MLDA}; see their Fig.~5. Their mean RV
near RJD~37557 is $-2.3$~\ks. As a~precaution, we do not use them in our
analyses.

File C: These near-UV/blue and yellow/red photographic spectra from Cerro
Tololo Observatory show the differences in the semi-amplitudes and
systemic velocities of RV curves of different spectral lines. For a better
comparison with our new spectra, we used the \ion{He}{i} RVs, but even so
one has to be aware that the RV zero-point might differ from other sources
due to line averaging and blending effects. The mean RV of the interstellar
\ion{Ca}{ii} lines for these spectra is $-1.6\pm1.2$~\ks. Although the
tabulated JDs are denoted as heliocentric ones by the authors, they are
regrettably only given to two decimal places (we note that the JDs and
HJDs start to differ only from the third decimal place onwards).

File D: These photographic spectra were secured at the coud\'e focus
of the ESO 1.5~m telescope in 1977-1978 and have a~linear dispersion
of 12~\AAMM. RVs were measured using the ARCTURUS oscilloscopic measuring
machine but the tabulated RVs are the mean RVs of components Aa1 and Ac1
for \ion{He}{i}~3820, 4026, and 4472~\AA, and \ion{Si}{iv}~4089~\AA\ lines.
Mean RVs of H$\gamma$, H8, H9, H10, and H11 are also tabulated.
The rms errors of the mean RVs are rather high, which is probably due to the effects
of line blending and differences in the RV amplitudes.
The RV zero-point of these mean RVs is therefore rather uncertain.
They do not cover the minimum RV of the 20\fd7 orbit. Also in this case,
Julian dates denoted as heliocentric ones are given to only two decimal places.

File E: These RVs are nine cross-correlation RVs from the SWP high-resolution
IUE images over the range 1250 to 1900~\AA. Their zero-point of the RV scale
was tuned via a mean interstellar velocity from three sources, and therefore
they seem usable for the monitoring of the systemic velocity changes.

File F: These RVs come from the first available electronic spectra secured
with the spectrograph Echelec attached to the ESO 1.5~m reflector in
1992 and 1993, and the coud\'e spectrograph CES attached to the
coud\'e auxiliary telescope (CAT) of the 1.4~m ESO reflector in 1994.
The Echelec spectra mainly cover the H$\beta$ and \ion{He}{i}~4922~\AA\ lines.
The CAT spectra contain only the neighbourhood of the
\ion{He}{i}~4922~\AA\ line and two of them were centred on \hae.
The authors warn that the H$\beta$ line is broad and less suitable
for RV measurements and that the 1992 spectra are affected by a flaw
between 4914 and 4917~\AA. Regrettably, we were unable to obtain
the original spectra and the RV zero-point is somewhat uncertain.

File H: These RVs come from the high-resolution Cantenbury University
HERCULES spectrograph and are based on a selection of symmetric,
mainly \ion{He}{i} lines but also some high-excitation lines.

\section{Details of the photometric data files and their homogenisation}
\label{apb}

\begin{table*}
\caption{All-sky mean \ubv\ magnitudes of the comparison and check stars
for \qz used by different observers in their differential photometries.}
\label{comps}
\begin{tabular}{ccccccclcc}
\hline\hline\noalign{\smallskip}
HD  & $V$ & $B$  & $U$ & \bv & \ub & Spectral&\ \ Source\\
    &(mag)&(mag)&(mag)&(mag)&(mag)&type\\
\hline\noalign{\smallskip}
    92740&    6.391&    6.482&    5.683&    $+0.091$&    $-0.799$&WN\,7&\citet{ma2010}\\
\color{blue}  93741&\color{blue} 7.23 &\color{blue} 7.22 &\color{blue} --   &\color{blue} $-0.01 $&\color{blue}         &B1\,II&\citet{cousins62}\\
    93131&    6.50 &    6.47 &    5.67 &    $-0.03$ &    $-0.80$ &WN\,6&\citet{walker72}\\
    93131&    6.496&    6.454&    5.572&    $-0.042$&    $-0.882$&WN\,6&Walker 2019     \\
    93131&    6.488&    6.477&    --   &    $-0.011$&     --     &WN\,6&\citet{black2020}\\
\color{blue} 93131&\color{blue} 6.484&\color{blue} 6.455&\color{blue} 5.546&\color{blue} $-0.029$&\color{blue} $-0.909$&WN\,6&\citet{ma2010}\\
    93131&    6.478&    6.448&    --   &    $-0.030$&     --     &WN\,6&Bohlsen Mirranook\\
\color{blue}  93222&\color{blue} 8.101&\color{blue} 8.166&\color{blue} 7.255&\color{blue} $+0.065$&\color{blue} $-0.911$&O7\,V&Zasche SAAO\\
    93222&    8.11 &    8.16 &    --   &    $+0.05$ &    --      &O7\,V&Bohlsen Mirranook\\
    93403&    7.265&    7.031&    --   &    $+0.234$&    --      &O5.5\,III&\citet{black2020}\\
    93683&    7.889&    7.761&    --   &    $+0.128$&    --      &B1e      &\citet{black2020}\\
\color{blue}  93695&\color{blue} 6.468&\color{blue} 6.343&\color{blue} 5.722&\color{blue} $-0.125$&\color{blue} $-0.621$&B5\,V&\citet{ma2010}\\
    93695&    6.427&    6.306&    --   &    $-0.121$&    --      &B5\,V&\citet{black2020}\\
    93695&    6.49 &    6.38 &    5.76 &    $-0.11$ &    $-0.62$ &B5\,V&Christie 2019\\
    93695&    6.47 &    6.34 &    --   &    $-0.13$ &    --      &B5\,V&Bohlsen Mirranook\\
    93843&    7.311&    7.290&    --   &    $-0.021$&    --      &O5\,III&\citet{black2020}\\
   305536&    8.94 &    8.94 &    --   &    $+0.00$ &    --      &O9.5\,V&Bohlsen Mirranook\\
\hline\noalign{\smallskip}
\end{tabular}
\tablefoot{Standard values, which we adopt here, are highlighted with
a~blue colour. HD 93683 = V736~Car was reported to be
a~17.8~d eclipsing binary.}
\end{table*}

We did our best to reduce the yellow, blue, and ultraviolet photometric
observations from various sources into a system that is similar to the
standard Johnson system. This is not an easy task because the observations
available to us consist of a mixture of differential observations measured
relative to various comparison and check stars, which are listed in
Table~\ref{comps}, and all-sky observations.

 As the three of the most frequently used comparisons were observed with
the TESS mission \citep{tess}, it is possible to provide some comments on
their photometric stability:
HD 93131 = TIC 390670978:
The TESS signals are mainly that of stochastic low-frequency variability.
The object is a Wolf Rayet (WR) star \citep[WN6ha-w;][]{wr1}. WR stars
are expected to exhibit stochastic low-frequency signals (likely due to
wind clumps), a good relatively large-amplitude example being WR40
studied by \citet{wr2}. The strongest signal in the frequency spectrum
of HR~93131 (which does not stand out above the red
noise `forest' of the low-frequency stochastic variability) is at
$f = 0.365697$~\cd, with an amplitude of 3.25~ppt. However, the maximum--minimum (max--min)
amplitude is 3.5\% (ten times higher than the strongest peak in the
frequency spectrum). As this variability is stochastic, it is not
expected to introduce any artificial signals in the target star when
differential photometry is performed. The main effect of this genuine
variability seen in TESS is an effective increase in the noise of the
ground-based measurements made relative to HD~93131. The standard deviation
of the relative flux measurements of TESS for this target is 0.0067.

\begin{figure}
\centering
\resizebox{\hsize}{!}{\includegraphics{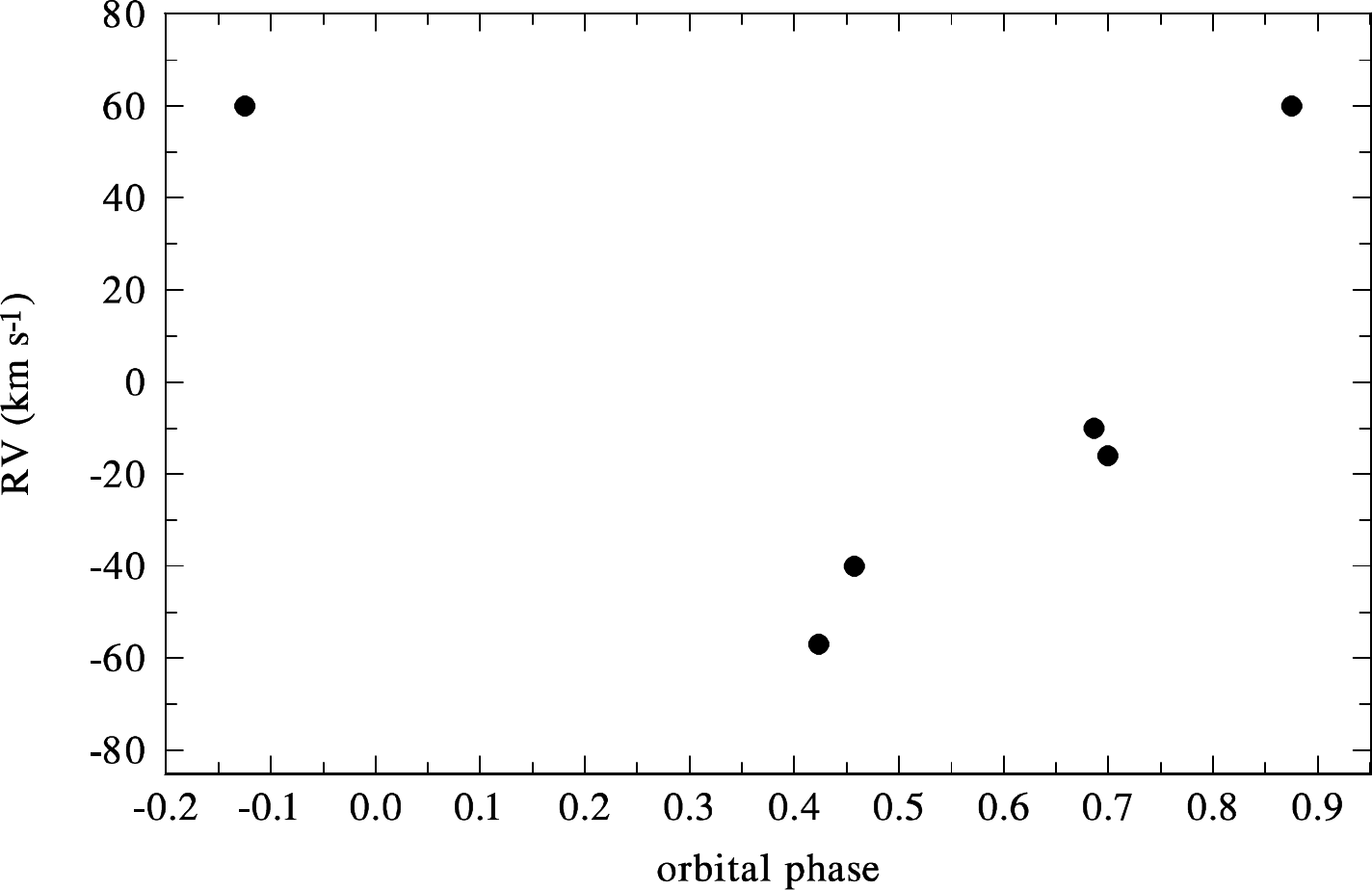}}
\resizebox{\hsize}{!}{\includegraphics{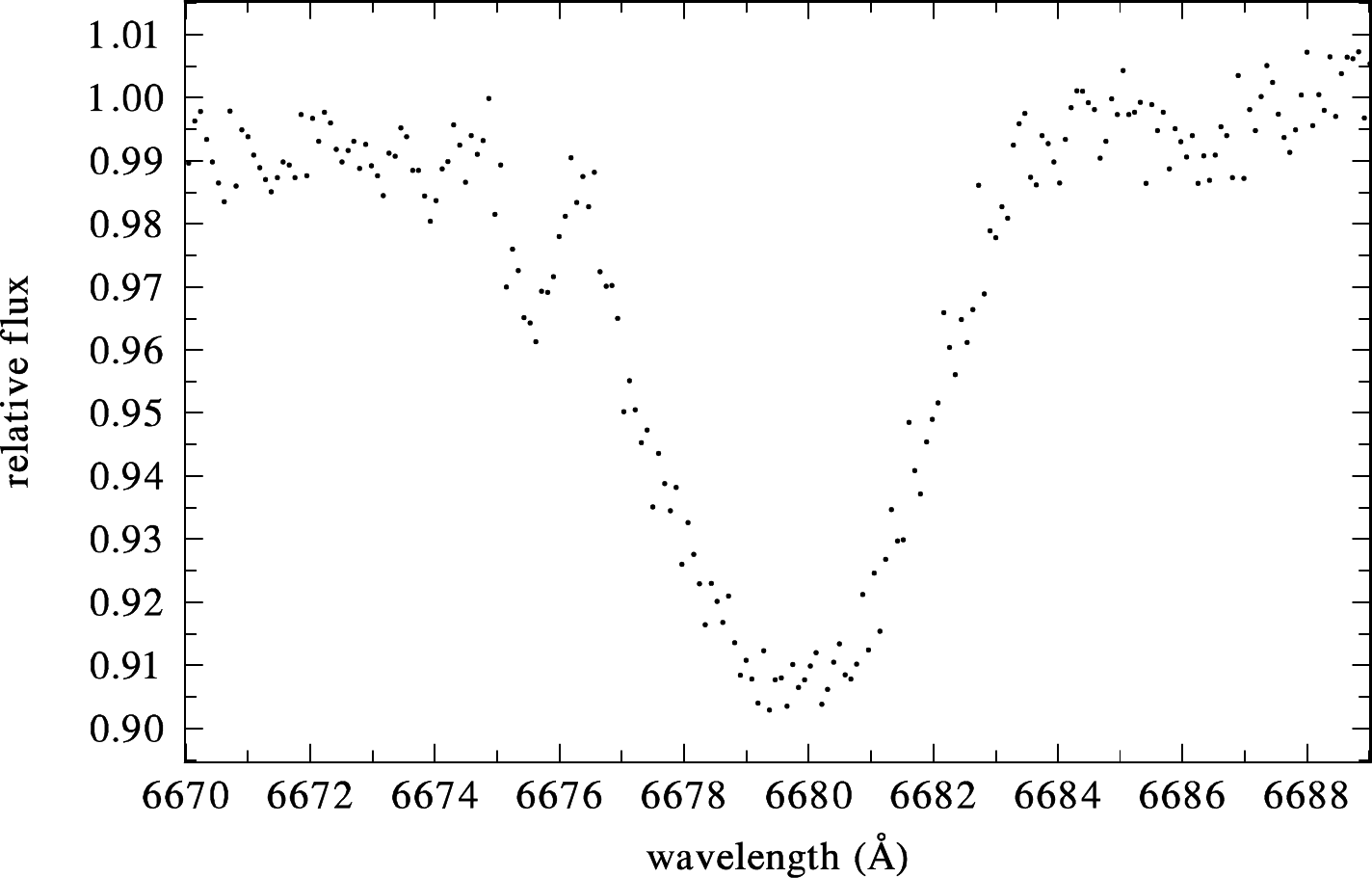}}
\caption{Top: Phase plot of \citet{class3} RVs of HD~93695 for the period
8\fd121289, which corresponds to the strongest $f_1$ frequency we find
in the TESS photometry. The epoch of phase zero is RJD~34095.37\,.
Bottom: Segment of one CHIRON spectrum of HD~93695 taken on November 26, 2021,
on which all \ion{He}{i} lines are split into one broad and one sharp
component. This proves that the object is SB2 composed of two B stars.}
\label{hd93695}
\end{figure}

 HD 93695 = TIC 459164314:
There are multiple low-amplitude periodic signals in the TESS data for this
system, plus some low-amplitude low-frequency stochastic excess. The
strongest five signals are
$f_1 = 0.123124$~\cd \ (amplitude = 0.83 ppt),
$f_2 = 0.247167$~\cd \ (clearly $2f_1$, amplitude = 0.77 ppt),
$f_3 = 0.56325$~\cd \ (amplitude = 0.65 ppt),
$f_4 = 1.291885$~\cd \ (amplitude = 0.2 ppt), and
$f_5 = 4.829406$~\cd \ (amplitude = 0.16 ppt). The max--min amplitude of
the TESS relative flux is 0.57\%, and the standard deviation is 0.0011.
Published spectral classifications of HD~93695 range from B3 \citep{class1}
over B5 \citep{class3} and B8/9 \citep{class2}. \citet{class3} also
reported rather large RV variations for this star. Figure~\ref{hd93695}
 shows a phase plot of the RVs of these latter authors versus 
 the phase of the 8\fd12189 period,
which corresponds to TESS frequency $f_1$. It is obvious that HD~93695
is a new spectroscopic binary, and probably has a slightly eccentric orbit.
As a free solution diverges to an unrealistically high eccentricity
because of an incomplete phase coverage, we derived only a trial circular
orbit solution, which gives $K = 65.6\pm8.0$~\ks, an epoch of maximum RV
of RJD~$34095.37\pm0.15$, and a systemic velocity of $+11.6\pm6.3$~\ks.
The first CHIRON CTIO echelle spectrum of the star was obtained on
November 26, 2021, and it clearly shows the presence of a sharp-lined B-type
secondary, which is seen in all \ion{He}{i} lines. A segment of this spectrum
in the neighbourhood of the \ion{He}{i}~6678~\AA\ line is shown in the bottom
panel of Fig.~\ref{hd93695}. A detailed study of this binary will be
the subject of a separate paper.
Obviously, the differential observations reduced relative to this
comparison should be treated with caution. They fortunately represent
only a small fraction of the data we are using here.

 HD 93222 = TIC 391050404:
As for HD 93131, the TESS variability appears to be purely that of
low-frequency stochastic excess. The strongest signal in the frequency
spectrum is at $f = 0.1672283$~\cd \ (amplitude = 0.8 ppt). The max--min
amplitude of the normalised TESS flux is 0.87\%, and the standard deviation
is 0.0014. Such low-amplitude stochastic variability is
typical for O stars viewed with TESS \citep{bowman2020}.
The star is listed as having a spectral type of O7V((f))z
by \citet{sota2014}.

Here, we provide details on the individual data sets
identified by their station numbers from Table~\ref{jouphot}.
Station 11:  These \ubv\ observations were obtained in 2004 by Marek Wolf
and in 2011 by Petr Zasche. They were transformed to the standard Johnson
system via non-linear transformations with the latest release of
the program HEC22 that is able to model time variability of the extinction during
the nights \citep{hhj94}.
Station 12: These \ubv\ observations were obtained by Pavel Mayer,
Horst Drechsel, and Reinald Lorentz and were later also re-reduced with
the program HEC22 to the standard system \citep{mayer92,ma2010}.
Station 61: These Hipparcos \hp\ observations were transformed to Johnson $V$
magnitude after \citet{hec98} assuming \bv = $-0$\m140, and \ub = $-0$\m832
for \qe. This seems justified because the \bv \ and \ub \ colours of \qz
show almost no variability with either the orbital phase of the Ac
system or with time. Only data with error flags 0 and 1 were used.
Station 93: These all-sky ASAS3 Johnson $V$ observations were processed in
several steps. First all observations of grade D  were omitted. As the data
are published for five different diaphragms, we selected the one with
the lowest average rms error, diaphragh 4 in our case. We then inspected
residuals with respect to the light curve and removed all deviating data points.
Station 100: These observations were obtained by Mark Blackford at
his backyard observatory in Chester Hill, Sydney, between 2015 and 2018.
A 0.080~m f/6 refractor and Canon 600D DSLR camera
were used. Typically, several hundred images were obtained each night and
5 images were always averaged. The exposure times were 20~s in 2015,
30~s in 2016 and 2017, and 50~s in 2018.
These data were re-reduced by M.B. in December 2020 relative
to HD~93131, with HD~93695 serving as the check star.
Station 105: These observations were secured by Anthony Moffat
\citep{moffat77} through a $\lambda5170$~\AA\ filter, 190~\AA\ wide.
We added the $V=8$\m101 magnitude of his comparison HD~93222 derived from
all-sky standard observations at Sutherland by Petr Zasche to obtain
approximate $V$ magnitudes of \qe. We assigned all these observations
with the rms error of 0\m007 estimated by the author. Moreover, we calculated
the heliocentric Julian dates from the published JDs.
Station 106: These discovery \ubv\ observations were published by
\citet{walker72}. Both the JDs and magnitudes were given
to only two decimal places there. A more extended data set was now
re-reduced by W.S.G.~Walker, corrected for extinction and transformed
to the Johnson system. Data between RJD~41290 and RJD~43270 come from
observations originally devoted to $\eta$~Car monitoring.
The data were observed in a sequence of three consecutive
10 s integrations, subsequently in $V$, $B$, and $U$, with a 2-5 s
delay between integrations in each filter. Julian dates were recorded
for the beginning of observations; therefore, to obtain mid-exposure
times of these observations, we increased Julian dates by 30+3+15
seconds, i.e. 0.00055~d. For observations with $V$ and $B$ filters only,
the respective correction was 0.00037~d. We then derived heliocentric
Julian dates for them. We also subtracted the \ubv\ values of the comparison
star HD~93131 derived by Walker in 2019, $V=$6\m496, \bv = $-0$\m042, and
\ub = $-0$\m882 and added the all-sky values of HD~93131
derived at La Silla instead (see Table~\ref{comps}).
Additional \ubv\ observations with the same instrument were obtained by
Grant Christie. A~part of these observations was used by \citet{minima98},
who incorrectly attributed them to the Mt John Observatory.
They were transformed to the standard Johnson system and reduced relative
to HD~93695, for which $V=$6\m49, \bv = $-0$\m11, and \ub = $-0$\m62 was
originally used. We re-reduced these observations relative to the standard
values of HD~93695 derived by \citet{ma2010} at LaSilla
(cf. again Table~\ref{comps}).
Station 107: The Bright Star Monitor Australia Station owned and operated by Peter Nelson.
This \bvri\ photometry was reduced by AH relative to a set
of comparison stars in the field of observations.
Station 108: Martin Bruce Berry Bright Star Monitor
\bvri\ photometry, also reduced by AH relative to a set
of comparison stars in the field of observations.
Station 109: These $BV$ observations were secured by an AAVSO observer
Terrence Bohlsen (AAVSO observer code BHQ) in Mirranook Observatory
and transformed to the standard Johnson system by him. They
consist of three subsets.
The observations in 2009 (RJDs 54945.05 - 55043.93) were secured with
an~SBIG ST9 camera. Later observations were secured with an~SBIG~ST10XME
camera. The observations from 2010 and two from January 2013
(RJD~55218.99 - 56324.01) were reduced differentially also relative
to HD~93131 while the 2013 to 2017 observations
(RJD~56324.01 - 57837.94) were reduced relative to an~ensemble of
three comparison stars, HD~93131, HD~93222, and HD~305536 and were
reported to AAVSO.
Station 110:
These observations were secured by the TESS space photometer \citep{tess}
in sectors 10 and 11 of the mission (between March 26, 2019, and May 21, 2019),
corresponding to 54.4 d of near-continuous observation.
Light curves were extracted from the
Full Frame Images (FFIs) at a~30-minute cadence for use in this work, where
the sky-background level was computed and subtracted from each frame by
JLB. This detrending method (sky-background subtraction) was preferred over
other detrending algorithms which can distort relatively high amplitude and
long-timescale variability (e.g. the 6 d binary orbital signal associated
with component Ac). Light curves for these sectors are also available with
two-minute cadence, which were used to search for periodic signals out to the
Nyquist limit of 360 \cd\ (after pre-whitening against the 6~d orbital
period and its many harmonics). Besides the 6~d orbital period and its
harmonics, no further periodic signals were found (although low-frequency
stochastic variability is present). A~neighbouring source, HD~305522, is
$66^{\arcsec}$ away from \qz and is approximately two~magnitudes fainter
than it is in the TESS band. We estimate that approximately 4\% of the flux in
the adopted aperture comes from this neighbour. However, a pixel-by-pixel
blending analysis demonstrates that the signals in the extracted light curve
can be confidently attributed to \qz and that the net effect of blending from
HD~305522 (and perhaps other much fainter neighbours) is a~very mild
suppression of the variability amplitudes.
Station 111: These observations were originally obtained as \ubv\ observations
by the late Harry Williams in his private Milton Road Observatory and reduced
by W.S.G.~Walker. However, for technical reasons (e.g. patina, which affected
one of the filter glasses) we could only use the $V$ band observations.
Stations 122 and 123: These $R$ band observations were secured with two
{\tt BRITE} nanosatellites \citep[see][]{weiss2014}, BRITE-Heweliusz
(BHr, station~122) in 2017, 2018, and 2020, and BRITE-Toronto (BTr,
station~123) in 2018 and 2019. The 2017 and 2018 observations have already
been used by \citet{black2020}. All data have now been reduced with the
latest version of the decorrelation software by A.P.
Observations in individual seasons are split into `setups' \citep[see][for
explanation of what setup refers to here]{popowicz2017}. Observations in a given
setup are reduced independently, which means that each setup has a
different zero-point in photometry, and magnitude shifts between setups
are allowed. However, there are exceptions. If the position of a subraster
was the same for different setups of the same satellite, two or more
setups were combined, applying magnitude offsets to minimise differences
between them. This is the case for station~122 data secured between
RJD~57784 and 57806, and station~123 data secured between RJD~58165 and 58187.
To reduce the scatter and remove possible systematic errors, we created
normal points averaged over the satellites' orbital periods of
0\fd068194 and 0\fd067431 for stations 122 and 123, respectively.
The number of these normal points is quoted in Table~\ref{jouphot}.
Station 124: These CCD $V$ band observations were obtained by Dave Blane
with a 150mm f/5 refractor and a Canon 1300D DSLR camera mounted on a~GEM
goto mounting. Ensemble aperture photometry was facilitated using the IRIS
software package. The camera $g$ magnitudes were linearly transformed to
the Johnson $V$ band using an MS-Excel spreadsheet.
Differential extinction was not applied. Each adopted measurement was the
average obtained from ten separate field images.
The systematic difference in the mean magnitudes for the two seasons can
probably be explained by the fact that different comparison star sets were
used. For the first season, the AAVSO values for six comparison stars were
used while for the second season, magnitude values for four comparison
stars recommended by Mark Blackford were used.
In the final step, the $V$ magnitude differences QZ~Car $-$ HD~93695
were added to our standard $V$ value for HD~93695 from Table~\ref{comps}.
Station 125: These observations were obtained by Mark Blackford at
Congarinni Observatory between 2019 and 2020. Slightly different
instrumentation was used in different seasons.
In 2020, an~80mm f/6 refractor and Atik One 6.0 CCD camera with Astrodon
Johnson $V$ filter were used to record images of \qe. Typically, 
about 40 images with two-second integration times were secured each night
(although the number and/or exposure time was different on a few nights). Instrumental
$V$ magnitude  differences \qz minus HD~93131 were derived, HD~93695 serving
as a check star. In 2019, the same instrumentation as in 2020 was used but
with a~2X tele extender. This meant that HD~93695
was outside the field of view, and so HD 305523 served as a~check star instead.
Several hundred images were usually obtained each observing night, with
10 second exposure times.
Station 126: These instrumental $V$ magnitude observations were obtained
by Martin Ma\v{s}ek relative to the comparison star HD~92741 with the FRAM
0.30~m automatic monitor at Pierre Auger Observatory in Argentina
\citep{fram2021} and kindly put at our disposal by their author. We added
$V$ =7\m23 to the magnitude differences var-comp.

\vfill\eject

\section{Supplementary tables}

\begin{table}[h]
\centering
\caption{Locally derived systemic velocities of system Aa.}
\label{gam-aa1}
\begin{tabular}{crrcl}
\hline\hline\noalign{\smallskip}
Mean & $\gamma$ \ \ \ \ \ \  &RVs &Range  & Notes\\
epoch& (\ks)        &No.&of RJDs& \\
\noalign{\smallskip}\hline\noalign{\smallskip}
35600& $-39.4\pm4.5$&  4& 35594.2--35653.2&uncertain \\
42821& $ -7.5\pm2.4$& 36& 42115.6--43240.6&  He I    \\
42795& $  6.3\pm2.7$& 28& 42117.7--43240.6&  Si IV   \\
42778& $ -6.6\pm3.0$& 27& 42117.7--43240.6&  H8-11   \\
43532& $  4.1\pm2.2$& 16& 43209.6--43534.9&  He/Si IV\\
43532& $ -4.7\pm3.8$& 16& 43209.6--43534.9&  H I     \\
44215& $  4.3\pm8.7$&  4& 43953.0--44376.2&  IUE     \\
49016& $-16.3\pm2.9$& 25& 48759.5--49148.6& \\
49451& $-21.4\pm3.4$&  5& 49448.6--49453.6& \\
54951& $  0.1\pm1.6$&  8& 53738.8--54956.6& \\
56205& $ -1.8\pm2.4$& 33& 55879.9--56459.6& \\
56984& $  0.7\pm2.8$& 26& 56668.0--57186.5& \\
58207& $  1.9\pm1.4$& 44& 58116.9--58262.5& \\
58883& $ -1.5\pm2.1$& 26& 58867.9--58909.7& \\
\hline
\end{tabular}
\end{table}

\begin{table*}[h!]
\centering
\caption{Locally derived systemic velocities of system Ac.}
\label{gam-ac1}
\begin{tabular}{crrcl}
\hline\hline\noalign{\smallskip}
Mean & $\gamma$ \ \ \ \ \ \  &RVs &Range  & Notes\\
epoch& (\ks)        &No.&of RJDs& \\
\noalign{\smallskip}\hline\noalign{\smallskip}
42891&$-27.7\pm 8.8$& 22& 42117.7--43238.5& He I\\
42880&$-30.1\pm15.3$& 15& 42117.7--43238.5& H I\\
43532&$-33.8\pm 9.1$& 16& 43209.6--43534.9& H5, H8-11\\
43532&$-46.6\pm 6.7$& 16& 43209.6--43534.9& He I, Si IV\\
44215&$-55.6\pm 6.0$&  4& 43953.0--44376.2& IUE        \\
49191&$ -1.3\pm 3.5$& 16& 48760.5--49813.3& Echelec, IUE, CAT\\
54951&$-29.4\pm 7.0$&  8& 53738.8--54956.6& FEROS, BESO  \\
56336&$-40.8\pm 5.4$& 18& 55880.9--56459.6& Hercules, BESO \\
57080&$-40.9\pm 6.6$& 20& 56670.0--57186.5& Hercules, CTIO \\
58206&$-29.9\pm 3.5$& 41& 58116.9--58262.5& CTIO         \\
58881&$-17.8\pm 6.6$& 22& 58867.9--58909.7& CTIO        \\
\hline
\end{tabular}
\end{table*}

\vfill\eject

\begin{table}
\begin{flushleft}
\caption{Instants of the primary minima derived from
the formal light-curve solutions of local data subsets with
the program \phoebee.}\label{minima}
\begin{tabular}{llrcrrrrrccc}
\hline\hline\noalign{\smallskip}
$T_{\rm min.I}$&error& No.& RJD range\\
(RJD)     &  (d)  \\
\noalign{\smallskip}\hline\noalign{\smallskip}
41632.8921&0.0082&  129&41035 -- 42234\\
42466.732 &0.035 &   34&42448 -- 42485\\
42772.536 &0.015 &   25&42583 -- 42962\\
43450.375 &0.020 &   20&43192 -- 43706\\
44308.243 &0.017 &   23&43915 -- 44699\\
44932.121 &0.017 &   24&44703 -- 45160\\
46425.694 &0.027 &   11&45340 -- 47513\\
48075.262 &0.016 &   46&47885 -- 48261\\
48429.266 &0.010 &   34&48287 -- 48571\\
48657.1438&0.0052&  214&48606 -- 48712\\
48861.1412&0.0064&   53&48750 -- 48971\\
49095.133 &0.017 &   76&49012 -- 49177\\
49413.095 &0.012 &  109&49372 -- 49459\\
52412.612 &0.015 &   37&51963 -- 52859\\
53054.4484&0.0050&  115&52925 -- 53190\\
53480.4469&0.0085&  113&53357 -- 53597\\
53780.3086&0.0072&  125&53653 -- 53909\\
54296.1959&0.0074&  127&54091 -- 54499\\
54650.1132&0.0090&  120&54502 -- 54801\\
54908.0536&0.0089&   97&54805 -- 55013\\
55297.955 &0.011 &   94&55014 -- 55579\\
56533.609 &0.017 &   41&56323 -- 56739\\
56971.5313&0.0073&  504&56811 -- 57191\\
57133.3705&0.0031& 1005&56973 -- 57288\\
57475.3391&0.0031&  939&57442 -- 57512\\
57781.2606&0.0024& 1401&57724 -- 57838\\
57889.2602&0.0026& 1987&57806 -- 57966\\
58213.1580&0.0027& 1776&58165 -- 58258\\
58285.1552&0.0043&  477&58259 -- 58315\\
58597.0780&0.0017& 3383&58512 -- 58681\\
58813.0076&0.0030&  716&58683 -- 58942\\
59010.8953&0.0035&  739&58942 -- 59075\\
\hline\noalign{\smallskip}
\end{tabular}
\tablefoot{See the text for details.}
\end{flushleft}
\end{table}

\begin{table}
\begin{flushleft}
\caption{Instants of the primary minima derived from the fits
of individual observed primary minima.}\label{pminima}
\begin{tabular}{ccrrrrrccc}
\hline\hline\noalign{\smallskip}
$T_{\rm min.I}$&error& \oc& epoch    \\
  (RJD)   &  (d)  &   (d) \\
\noalign{\smallskip}\hline\noalign{\smallskip}
41213.0394& 0.0612&  0.1182&-2214.0\\
41464.9476& 0.0155&  0.0816&-2172.0\\
41986.7909& 0.0176&  0.0393&-2085.0\\
42466.7372& 0.0145&  0.0908&-2005.0\\
44164.2203& 0.0165& -0.0541&-1722.0\\
47217.4360& 0.0657& -0.1692&-1213.0\\
48273.2238& 0.0106& -0.1501&-1037.0\\
48687.1382& 0.0015& -0.1450& -968.0\\
48687.1402& 0.0042& -0.1430& -968.0\\
48867.1411& 0.0088& -0.1026& -938.0\\
49401.0168& 0.0849& -0.1099& -849.0\\
52820.5406& 0.0030&  0.1632& -279.0\\
53474.3981& 0.0659&  0.1640& -170.0\\
53780.3261& 0.0227&  0.1591& -119.0\\
54242.2179& 0.0161&  0.1521&  -42.0\\
54554.1235& 0.0057&  0.1260&   10.0\\
54956.0241& 0.0164&  0.1147&   77.0\\
55147.9468& 0.0588&  0.0795&  109.0\\
55297.9183& 0.0162&  0.0839&  134.0\\
56905.5194& 0.0064&  0.0373&  402.0\\
57079.4823& 0.0192&  0.0383&  431.0\\
57121.4978& 0.0281&  0.0630&  438.0\\
57133.4239& 0.0151& -0.0083&  440.0\\
58573.0641& 0.0010& -0.0525&  680.0\\
58579.0753& 0.0010& -0.0400&  681.0\\
58585.0711& 0.0010& -0.0429&  682.0\\
58591.0669& 0.0010& -0.0458&  683.0\\
58603.0611& 0.0010& -0.0490&  685.0\\
58609.0362& 0.0010& -0.0726&  686.0\\
58615.0690& 0.0010& -0.0384&  687.0\\
58621.0322& 0.0010& -0.0739&  688.0\\
58621.0747& 0.0125& -0.0314&  688.0\\
59046.9775& 0.0157& -0.0353&  759.0\\
\hline\noalign{\smallskip}
\end{tabular}
\tablefoot{The residuals were derived for a~mean period of 5\fd99868537
and epoch RJD~54494.0106. See the text for details.}
\end{flushleft}
\end{table}

\begin{table}
\begin{flushleft}
\caption{Instants of the secondary minima derived from the fits
of individual observed primary minima.
}\label{sminima}
\begin{tabular}{ccrrrccc}
\hline\hline\noalign{\smallskip}
$T_{\rm min.I}$&rms& \oc& epoch    \\
  (RJD)   &  (d)  &   (d) \\
\noalign{\smallskip}\hline\noalign{\smallskip}
41216.0479& 0.0375& 0.1273&-2213.5\\
41467.9433& 0.0157& 0.0779&-2171.5\\
41989.7736& 0.0180& 0.0226&-2084.5\\
42469.6886& 0.0582& 0.0428&-2004.5\\
44167.2250& 0.0111&-0.0488&-1721.5\\
48276.2083& 0.0435&-0.1650&-1036.5\\
48690.1339& 0.0054&-0.1486& -967.5\\
48690.1660& 0.0036&-0.1165& -967.5\\
48870.0942& 0.0193&-0.1489& -937.5\\
49404.0753& 0.0430&-0.0508& -848.5\\
52823.5305& 0.0157& 0.1537& -278.5\\
53477.4037& 0.0314& 0.1702& -169.5\\
53783.3012& 0.0133& 0.1348& -118.5\\
54245.2047& 0.0174& 0.1395&  -41.5\\
54557.1457& 0.0177& 0.1489&   10.5\\
54959.0411& 0.0202& 0.1323&   77.5\\
55150.9995& 0.0828& 0.1328&  109.5\\
55300.9505& 0.1260& 0.1167&  134.5\\
56908.5229& 0.0140& 0.0414&  402.5\\
57082.5011& 0.0043& 0.0577&  431.5\\
57124.4981& 0.0456& 0.0639&  438.5\\
57136.4889& 0.0648& 0.0573&  440.5\\
58570.0600& 0.0010&-0.0573&  679.5\\
58576.0796& 0.0010&-0.0364&  680.5\\
58588.0595& 0.0010&-0.0539&  682.5\\
58594.0815& 0.0010&-0.0306&  683.5\\
58600.0523& 0.0010&-0.0585&  684.5\\
58606.0506& 0.0010&-0.0589&  685.5\\
58612.0606& 0.0010&-0.0475&  686.5\\
58618.0747& 0.0010&-0.0321&  687.5\\
58624.0673& 0.0118&-0.0382&  688.5\\
\hline\noalign{\smallskip}
\end{tabular}
\tablefoot{The residuals were derived for a mean period of 5\fd99868537
and epoch RJD~54491.0113. See the text for details.}
\end{flushleft}
\end{table}

\end{appendix}
\end{document}